\documentstyle[emulateapj,psfig]{article}

\begin{document}
 
\title{Non-LTE Models and Theoretical Spectra of Accretion Disks
in Active Galactic Nuclei.  III.  Integrated Spectra for Hydrogen-Helium
Disks}
 
\author{Ivan Hubeny}
\affil{AURA/NOAO, NASA Goddard Space Flight Center, Code 681, Greenbelt, 
       MD 20771}

\author{Eric Agol}
\affil{Department of Physics and Astronomy, Johns Hopkins University,
           Baltimore, MD 21218 }

\author{Omer Blaes}
\affil{Department of Physics, University of California, Santa Barbara,
CA 93106}

\and
\author{Julian H. Krolik}
 
\affil{Department of Physics and Astronomy, Johns Hopkins University,
           Baltimore, MD 21218}
 
\begin{abstract}
We have constructed a grid of non-LTE disk models for a wide
range of black hole mass and mass accretion rate,
for several values of the viscosity parameter $\alpha$,
and for two extreme values of the  black hole spin: the maximum-rotation
Kerr black hole, and the Schwarzschild (non-rotating) black hole.
Our procedure calculates self-consistently the vertical
structure of all disk annuli together with the radiation field, without
any approximations imposed on the optical thickness of the
disk, and without any ad hoc approximations to the behavior of
the radiation intensity.
The total spectrum of a disk is computed by summing the
spectra of the individual annuli, taking into account the
general relativistic transfer function.

The grid covers nine values of the black hole mass between
$M = 1/8 \times 10^9$ and $32 \times 10^9$~M$_\odot$ with a two-fold 
increase of mass for each subsequent value; and eleven values of the mass
accretion rate, each a power
of 2 times 1~$M_\odot$~yr$^{-1}$.  The highest value of the accretion rate
corresponds to the total luminosity $L/L_{\rm Edd} \approx 0.3$.
We show the vertical structure of individual annuli within
the set of accretion disk models, along with their local emergent flux,
and discuss the internal physical self-consistency of the models.
We then present the full disk-integrated spectra, and discuss
a number of observationally interesting properties of the models, such as
optical/ultraviolet colors, the behavior of the hydrogen Lyman limit region,
polarization, and the number of ionizing photons.    Our calculations
are far from definitive in terms of the input physics, but generally we
find that our models exhibit rather red optical/UV colors.  Flux
discontinuities in the region of the hydrogen Lyman limit are only present
in cool, low luminosity models, while hotter models exhibit blueshifted
changes in spectral slope.
\end{abstract}


 
\section{Introduction}

A black hole that steadily accretes gas from its surroundings at high rates,
but below the Eddington limit, should form a geometrically thin accretion disk
supported by the residual angular momentum of the gas.  For black hole masses
in the range $10^8-10^9$~M$_\odot$, thought to be typical of bright active
galactic nuclei (AGN) and quasars, the peak effective temperature is
expected to be around $10^4 - 10^5$ K.  This estimate
is roughly consistent with the observation that quasar spectra peak
in the ultraviolet, bolstering the belief that quasars are indeed powered
by accretion onto massive black holes.

Over the years, many authors have attempted to calculate detailed theoretical
spectra of geometrically thin, optically thick accretion disks in order to
compare with observation, with varying degrees of sophistication (e.g.
Kolykhalov \& Sunyaev 1984; Sun \& Malkan 1989; Laor \& Netzer 1989; Ross,
Fabian, \& Mineshige 1992; Shimura \& Takahara 1993, 1995; D\"orrer et
al. 1996; and Sincell \& Krolik 1998).  As reviewed recently by Krolik (1999a)
and Koratkar \& Blaes (1999), these and other models generally suffer
a number of problems when trying to simultaneously explain various features
of the observations, e.g. optical/ultraviolet continuum
spectral shapes, the lack of observed features at the Lyman limit of hydrogen,
polarization, the origin of extreme ultraviolet and X-ray emission, and
correlated broadband variability.  It may be that the resolution of these
problems requires drastic revision of the accretion disk paradigm.
Alternatively, it may simply be that the theoretical modeling to date is
still too crude to do justice to the inherent complexities of the accretion
flow.  In particular, inclusion of non-LTE effects and detailed opacity
sources, Comptonization, and interaction between the disk and X-ray producing
regions should all be taken into account.

We have embarked on a long term program to construct detailed
model spectra of accretion disks in the axisymmetric, time-steady, thin-disk 
approximation.  At the peak temperatures, the most important opacity is 
provided by electron scattering, but bound-free and free-free continuum
opacities due to hydrogen and helium can also be significant, so we include 
these opacities in this study.  Since scattering can be the dominant 
opacity and the densities can be rather low, departures from local 
thermodynamic equilibrium (LTE) can be significant, so we include non-LTE 
effects as well.  We include the effects of relativity on the disk structure 
and on the transport of radiation from the disk to infinity.  We construct 
a grid of models for a black hole spin of $a/M=0$ (Schwarzschild black
hole), and 0.998 (maximum rotation Kerr black hole), luminosities between 
$3\times10^{-4}\, L_{\rm Edd}$ and $0.3\, L_{\rm Edd}$, and black hole masses
between $M_9\equiv10^{-9} M/M_\odot = 0.125$ to $M_9=32$.  To determine
the surface mass 
density of the disk, we assume that the viscous stress scales as $t_{r\phi} 
= \alpha P_{\rm total}$ (Shakura \& Sunyaev 1973), where $t_{r\phi}$ and
$P_{\rm total}$ are the vertically integrated viscous stress and total
vertically-integrated pressure, respectively.  

In previous papers (Hubeny \& Hubeny 1997 and 1998a, hereafter Papers I and II
respectively), we presented detailed spectra and the vertical structure
of individual annuli.  In this paper we integrate such spectra over radius 
for a grid of 99 mass/accretion rate combinations appropriate for quasars.  
We do not include annuli with low effective temperatures 
($T_{\rm eff} < 4000$ K) 
as these require molecular opacities for accurate computation.  The spectra 
of hot annuli can be affected by Compton scattering, which we have not 
included in the calculation, but will in future work.  It is likely that
metal opacities will also modify the final spectrum (cf. Hubeny \& Hubeny
1998b), so we consider this work as a benchmark for future metal line-blanketed
models.

This paper is organized as follows.  In section 2 we describe
our model assumptions and computational methods.  Then in section 3, which
constitutes the bulk of the paper, we present our results.  We start by
showing the vertical structure of individual annuli within
the set of accretion disk models, along with their local emergent flux.
We then discuss the internal physical self-consistency of these models,
before presenting the full disk-integrated spectra.  We finish section 3
with a discussion of a number of observationally driven issues:
optical/ultraviolet colors, spectra in the hydrogen Lyman limit region,
polarization, and ionizing continua.  Finally, in section 4 we summarize
our conclusions, in particular pointing out the additional physics which
will be included in future papers of this series.

\section{Model Assumptions and Computations}

To construct a model of an accretion disk, we assume the vertical
disk structure can be well approximated by one-dimensional equations;
that is, we assume the disk is locally plane parallel.
We assume that, on average, the disk
is static in the corotating frame (in reality disks are subject to
many instabilities which invalidate this approximation), and that the
only energy transport is due to radiation flux in the vertical direction,
i.e. we ignore convection and conduction.
By assuming time-steadiness and local radiation of dissipated
heat, we can write down and solve the equations for the disk
structure (Page \& Thorne 1974); these equations are summarized in Paper II. 
For a given radius $r$, the (one-sided) flux (and thus the effective
temperature, $T_{\rm eff}$) is determined by
\begin{equation}
\label{teff}
F(r) \equiv \sigma_B T_{\rm eff}^4 = {3 G M \dot M\over 8 \pi r^3}R_R,
\end{equation}
where $\sigma_B$ is the Stefan-Boltzmann constant, $\dot M$ the mass
accretion rate, and $R_R$ is a relativistic correction factor 
(Page \& Thorne 1974, in the notation of Krolik 1999a).  
Following the usual practice for
geometrically thin accretion disks, we assume that there is no torque at
the innermost stable circular orbit in all our disk models.  There
are reasons to question this assumption (Krolik 1999b,
Gammie 1999); if it fails, the disk spectrum and polarization could
be substantially changed (Agol \& Krolik 1999).

We have calculated models for two values of the viscosity parameter:
$\alpha = 0.01$, and $\alpha = 0.1$.
The choice of $\alpha = 0.01$ is near the
value expected from simulations of the magneto-rotational instability
in accretion disks (e.g. Balbus \& Hawley 1998), while $\alpha = 0.1$
represents a typical value of the viscosity parameter used in other studies.
Smaller $\alpha$ or a stress which scales in
proportion to the gas pressure would lead to a larger surface
mass density, higher density, and, presumably, spectra closer to LTE.

If the disk rotates on cylinders, and the angular frequency of each
cylinder is the one appropriate to a circular orbit at the midplane,
there is a vertical component to the effective gravity proportional to
height $z$ above the disk midplane.  We ignore 
the self-gravity of the disk, an excellent approximation for the radii
important to our problem.

We have assumed that the local dissipation is proportional
to the local density, except for the top 1\% of the disk where we force
the dissipation to decline -- see Paper II.  This distribution is chosen
in order to yield a hydrostatic equilibrium solution in the bulk of the
disk when radiation dominates (cf. Shakura \& Sunyaev 1973) while retaining
the possibility of thermal balance in the outer-most layers even in the
absence of Comptonization.
In Paper II (in particular, 
see figures 11 and 12 there) we showed that neither the choice of the division 
point between the regions of vertically constant and declining viscosity,
nor the slope of the power law for the viscosity in the declining regime, 
changes the predicted continuum emergent spectrum significantly.
Disks with this structure can be convectively
unstable, however.  Convection can be expected to accelerate heat loss,
leading to a disk structure that is rather thinner and denser (Bisnovatyi-Kogan
\& Blinnikov 1977), although perhaps not substantially so (Shakura, Sunyaev
\& Zilitinkevich 1978).  
Even annuli with entropy increasing upward, so that they are
stable according to the usual Schwarzschild criterion, may still nevertheless
be subject to convective instabilities mediated by thermal conduction along
weak magnetic field lines (Balbus 1999).  How this would manifest itself in
the presence of turbulence driving the radial angular momentum transport is
not clear, however.  In any case, as stated above, we completely ignore
convective heat transport in all our models here.

When either radiation or gas pressure dominates, the surface
mass density may be readily calculated in the grey, one-zone, diffusion
approximation.  However, when the two are comparable to each other, 
the disk structure equations combine to form a single tenth-order
polynomial equation in one of the variables (e.g., $\Sigma^{1/4}$).
We solve this equation iteratively, using the two limiting regimes to
bracket the root.

Once the effective temperature, surface mass density, and dissipation
profile are determined, the vertical disk structure can be computed.
To compute the vertical structure of a given annulus, we solve
simultaneously the entire set of structural equations: hydrostatic equilibrium 
in the vertical direction; local energy balance; radiative transfer;
and, since we are not
generally assuming LTE, statistical equilibrium for all selected
energy levels of all selected atoms and ions.  The equations are solved
for the whole extent of the disk between the midplane and the surface
using proper boundary conditions.  For details, the reader is referred
to Paper II.  We stress that no ad hoc assumptions about the nature of
the radiation field or the radiative transfer are made; for instance, we do not
use the diffusion approximation or an escape probability treatment,
or any assumption about the angular dependence of the specific
intensity; the radiative transfer is solved exactly.  To represent
the radiation field, we use about 150 frequency points placed to define
all continuum edges and resolve any frequency-dependent structure in
the emergent intensity.  The minimum frequency is set to $10^{12}$ Hz,
while the highest frequency is chosen so that even at the midplane
the intensity for frequencies higher than the maximum is negligible.
In terms of the midplane temperature (Paper II and Hubeny 1990)
\begin{equation}
\label{tmid}
T_{\rm mid} \approx T_{\rm eff} (3 \tau_{\rm mid}/8)^{1/4} = 
T_{\rm eff} [(3/8) (\Sigma/2) (\chi_R)]^{1/4} \, ,
\end{equation}
this goal can be achieved by setting the maximum frequency to
\begin{equation}
\nu_{\rm max} = 17 kT_{\rm mid}/h .
\end{equation}
Here $\chi_R$ is the
Rosseland mean opacity (per gram), which we take for simplicity to be 0.34,
the value corresponding to an opacity dominated by electron scattering
in a fully ionized H-He plasma of solar abundance.

We consider disks composed of hydrogen and helium only.
We include metals in computing the molecular weight, but for this work, 
we ignore their effects on the opacity.
Hydrogen is represented essentially exactly: the first 8 principal
quantum numbers are treated separately, while the upper levels are
merged into a single non-LTE level accounting for level dissolution
as described by Hubeny, Hummer, \& Lanz (1994).  We do, however,
assume complete $l$-mixing; given the high electron density in these
environments, this should be a good approximation.  Neutral
helium is represented
by a 14-level model atom, which incorporates all singlet and triplet
levels up to $n=8$. The 5 lowest levels are included individually;
singlet and triplet levels are grouped separately from $n=3$ to $n=5$,
and we have formed three superlevels for $n=6, 7,$ and~8.
The first 14 levels of He$^+$ are explicitly treated. We assume a solar
helium abundance, $N({\rm He})/N({\rm H}) = 0.1$.

The opacity sources we include are all bound-free transitions (continua)
from all explicit levels of H, He I, and He II; free-free transitions
for all three ions, and electron scattering.  For the coolest models
($T_{\rm eff} < 9000$ K), we also consider the H$^-$ bound-free and 
free-free opacity, assuming LTE for the H$^-$ number density.
In this paper, we 
assume coherent (Thompson) scattering.  As was discussed in Paper II,
effects of non-coherent (Compton) scattering are negligible for
models with $T_{\rm eff}$ around or below $10^5$~K. Therefore, most
models of the present grid (see Sect.3) are not influenced by the
effects of Comptonization, although the hottest ones may be. We have recently
implemented Comptonization in our modeling code, and checked that
this is indeed the case; however, we choose to neglect Comptonization
in this paper in order to provide a benchmark grid of models computed
using classical approximations of H-He composition and without
Comptonization.  The effects of Compton scattering will be included in
a future paper, where we will also discuss in detail its influence on
the emergent spectra.  It can be expected to become especially important
when the dissipation per unit mass is enhanced near the disk's surface.

The structure equations are highly nonlinear, but are very similar
to corresponding equations for model stellar atmospheres. 
We use here the computer program TLUSDISK, which is a derivative of the
stellar atmosphere program TLUSTY (Hubeny 1988). The program
is based on the hybrid complete-linearization/accelerated lambda 
iteration (CL/ALI) method (Hubeny \& Lanz 1995). The method resembles 
traditional complete linearization, however the radiation intensity 
in most (but not necessarily all) frequencies
is not linearized; instead it is treated via the ALI scheme (for a
review of the ALI method, see e.g., Hubeny 1992). Moreover, we use
Ng acceleration and the Kantorovich scheme (Hubeny \& Lanz 1992)
to speed the solution.
We start with a grey atmosphere solution (Hubeny 1990), first solving for 
the disk structure assuming that the statistical
equilibrium is described by LTE.  Using this as a starting point, we drop the 
LTE assumption and compute the disk structure using the full statistical 
equilibrium equations, as described in Paper II, but assuming that the
line transitions are in detailed radiative balance.  In other words,
the statistical equilibrium equations explicitly contain the collisional
rates in all transitions, and radiative rates only in the continuum
transitions.
We have considered 70 discretized depth points.  The
top point is set to $m_1=10^{-3}$ g cm$^{-2}$, where $m$ is the column mass,
i.e., the mass in a column above a given height.  The last depth point
is the column mass corresponding to the midplane, and is given by $\Sigma/2$.
The depth points are equally spaced in logarithm between these two values.
The models were computed on a DEC Alpha with 500 MHz clock speed; 
with the above values for the number of frequency and depth points, the LTE
models for individual rings required typically 5-10 iterations with
approximately 2.5 seconds
per iteration, while the non-LTE models required typically 5-10 (for hotter
annuli), or 10-30 iterations (for cooler annuli), with about 6 seconds 
per iteration.

We do not compute here models treating radiative rates
in line transitions explicitly. We have considered such models in Paper II
and found that including lines explicitly does not change the vertical structure
or emergent continuum radiation significantly. These models are also
much more time consuming. However, the most important physical point is
that we have found in test calculations that line profiles are
influenced significantly by the effects of Compton scattering.  Since we
are neglecting the Comptonization here to provide a benchmark grid of
classical H-He models without Comptonization, we feel that
including lines at this stage would not be much more than a numerical
exercise.  We therefore defer treatment of more realistic models
including lines and Comptonization to a future paper.

In the course of calculating atmosphere models, we sometimes ran into
difficulties with convergence.
For very low and very high temperatures, we could not get convergence,
a problem which may be ameliorated in the future by including more sources 
of opacity (such as metals at high temperatures and molecules at low 
temperatures).  At certain radii and certain depths within the disk, we 
found ionization fronts in helium around $T_{\rm eff} \sim 35,000$K which are 
very narrow in extent and very sensitive to 
the temperature within the disk, causing limit cycle behavior where the
helium alternates between being mostly doubly ionized or recombined 
to a singly ionized stage during
successive steps, preventing solution of the atmosphere structure.  We solved
this problem by computing the disk structure for radii just smaller or larger 
than the radii where the front exists, and then using these solutions as 
starting solutions for the radii at which the fronts exist.  In practice,
the range of radii where this problem exists is narrow so that the uncertainty
in the structure does not affect the overall disk spectrum.  We also found
similar He I/He II ionization fronts at lower temperatures, $T_{\rm eff} \sim
15,000$K, and for hydrogen around and below 9000 K.

To find the total disk spectrum, we divided the disk into 25-35 radial rings,
spaced (roughly) logarithmically.  At each ring, after computing the vertical 
structure, we perform a detailed radiation transfer solution for 
the Stokes vector as a function of frequency and angle.
The spectrum is found by integrating the total emergent intensity
over the disk surface using our relativistic transfer function code (Agol
1997).  
The transfer function computes the trajectories of photons
from infinity to the disk plane, finding the emitted radius, redshift,
and intensity at each image position at infinity for a given observation
(Cunningham 1975).  In this paper,
we neglect the effects of radiation which returns to the
accretion disk.

\section{Results}

The parameter space of our grid of models is displayed in figure \ref{FIGGRID}.
The defining parameters are $M_9$, the black hole mass expressed in
$10^9 M_\odot$, and $\dot M$, the accretion rate in units of
$M_\odot$ yr$^{-1}$.
The grid covers nine values of the black hole mass between
$M_9$ = 1/8 and 32; each subsequent mass is twice the previous mass.
An analogous approach is used for the mass accretion rate, i.e. eleven values
for each black hole mass which are powers
of 2 times 1~$M_\odot$~yr$^{-1}$.  The highest value of the accretion rate
in each case is chosen to make $L/L_{\rm Edd} =0.286$;
i.e., $\dot M (M_\odot\, {\rm yr}^{-1}) = 2 M_9$.  In the following text,
we refer to this highest value as $L/L_{\rm Edd} =0.3$.
the ten subsequent values
have half the accretion rate (and luminosity) of the previous model.
Our grid spans a range similar to models which have been previously used to
fit quasar spectra (Sun \& Malkan 1989; Laor 1990).

Our basic grid assumes $a/M = 0.998$.  However, we have also
constructed a parallel grid for a Schwarzschild black hole with
the same black hole masses and values of
$L/L_{\rm Edd}$.  The mass accretion rates for $a/M = 0$ are a factor of
5.613 higher than the corresponding values for $a/M = 0.998$ because
the radiative efficiency of a Kerr disk is that much greater (e.g. Shapiro
\& Teukolsky 1983).  

\vskip 2mm
\hbox{~}
\centerline{\psfig{file=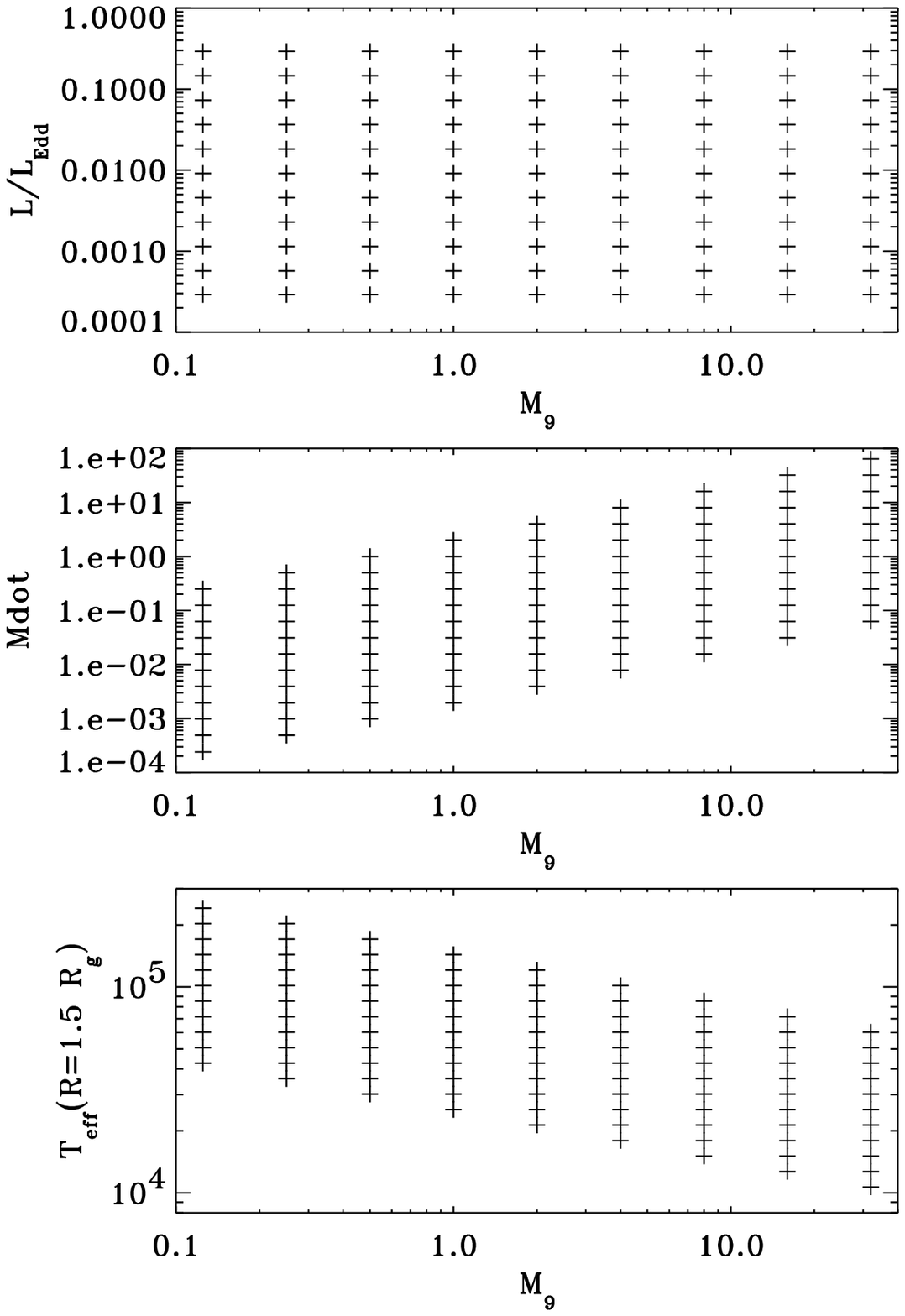,width=3.6in}} 
\noindent{
\scriptsize \addtolength{\baselineskip}{-3pt}
\vskip 1mm
\figcaption{
The parameters of our grid of disk models for
the Kerr hole case.
Upper panel: $L/L_{\rm Edd}$ versus black hole mass,
expressed in $10^9 M_\odot$. Middle panel: the mass accretion rate, $\dot M$,
expressed in $M_\odot$ yr$^{-1}$.
Lower panel: the maximum effective temperature (which is found at
$r \simeq 1.5\, r_g$). \label{FIGGRID}}
\vskip 3mm
\addtolength{\baselineskip}{3pt}
}

For all disk models, we compute detailed vertical structure at the
following radii (expressed as $r/r_g$, where $r_g = GM/c^2$ is the
gravitational radius):
$r/r_g$ = 1.5, 1.7, 2, 2.5, 3, 3.5, 4, 5, 6, 7, 8, 9, 10, 
12, 14, 16, 18, 20, 25, 30, 40, 50, 60, 70, 80, 90, 100,
120, 140, 160, 180, 200, 250, 300, 400, 500; 
that is, while the corresponding effective temperature $T_{\rm eff} >
4000$ K.  For the Schwarzschild grid, we use the above values of
$r/r_g$, multiplied by 5.
The actual number of computed annuli thus depends on the basic
parameters of the disk.

\subsection{Properties of Models}

In this section, we discuss the behavior of individual annuli, while in the
rest of the paper we will present the emergent radiation integrated over the
whole disk.
We have chosen a model with  $M_9 = 1$, $\dot M=1$ (i.e., $L/L_{\rm Edd}=0.15$)
as a representative model for displaying various quantities. The
behavior of other individual disk models is similar.
Figure \ref{FIGTEMP} displays the local electron temperature as a function
of position for all individual annuli. The position is expressed as 
column mass, $m$, above the given depth.  As mentioned in Sect.~2, the
uppermost point was chosen
to be $m= 10^{-3}$ g cm$^{-2}$, while the highest value corresponds to the
midplane of the disk. The temperature is a nearly monotonic function of depth,
although there is a slight temperature rise at the surface for some
models. The detailed behavior of temperature for several representative annuli
was discussed extensively in Paper II.
The behavior of temperature for all annuli is easily understood.
In the LTE approximation, the midplane temperature is given
by equation (\ref{tmid}), i.e. it is
proportional to $T_{\rm eff}(r) [m_0(r)]^{1/4}$, where
$m_0 \equiv \Sigma/2$
is the total column density at the midplane. Neglecting for simplicity
the relativistic corrections, $T_{\rm eff} \propto r^{-3/4}$ - see equation
(\ref{teff}), while the column density $m_0 \propto r^{3/2}$ for radiation
pressure dominated annuli (see Eq. 18 of Paper II), which is the case for
the models considered in figure \ref{FIGTEMP}.
Therefore, the LTE approximation predicts that the
midplane temperature, $T_{\rm mid} \propto r^{-3/8}$.
In contrast, the surface temperature is proportional to $T_{\rm eff}$;
therefore, $T_{\rm surf} \propto r^{-3/4}$.
Figure \ref{FIGTEMP} shows that these scalings do in fact hold approximately.
The ratio of the lowest and highest radii is roughly 45, so if the
LTE approximation held, the surface
temperature would vary by a factor of 17 at the surface, and by a factor of
4.2 at the midplane.  The model values are 14 and 4.6, respectively.

%
\vskip 2mm
\hbox{~}
\centerline{\psfig{file=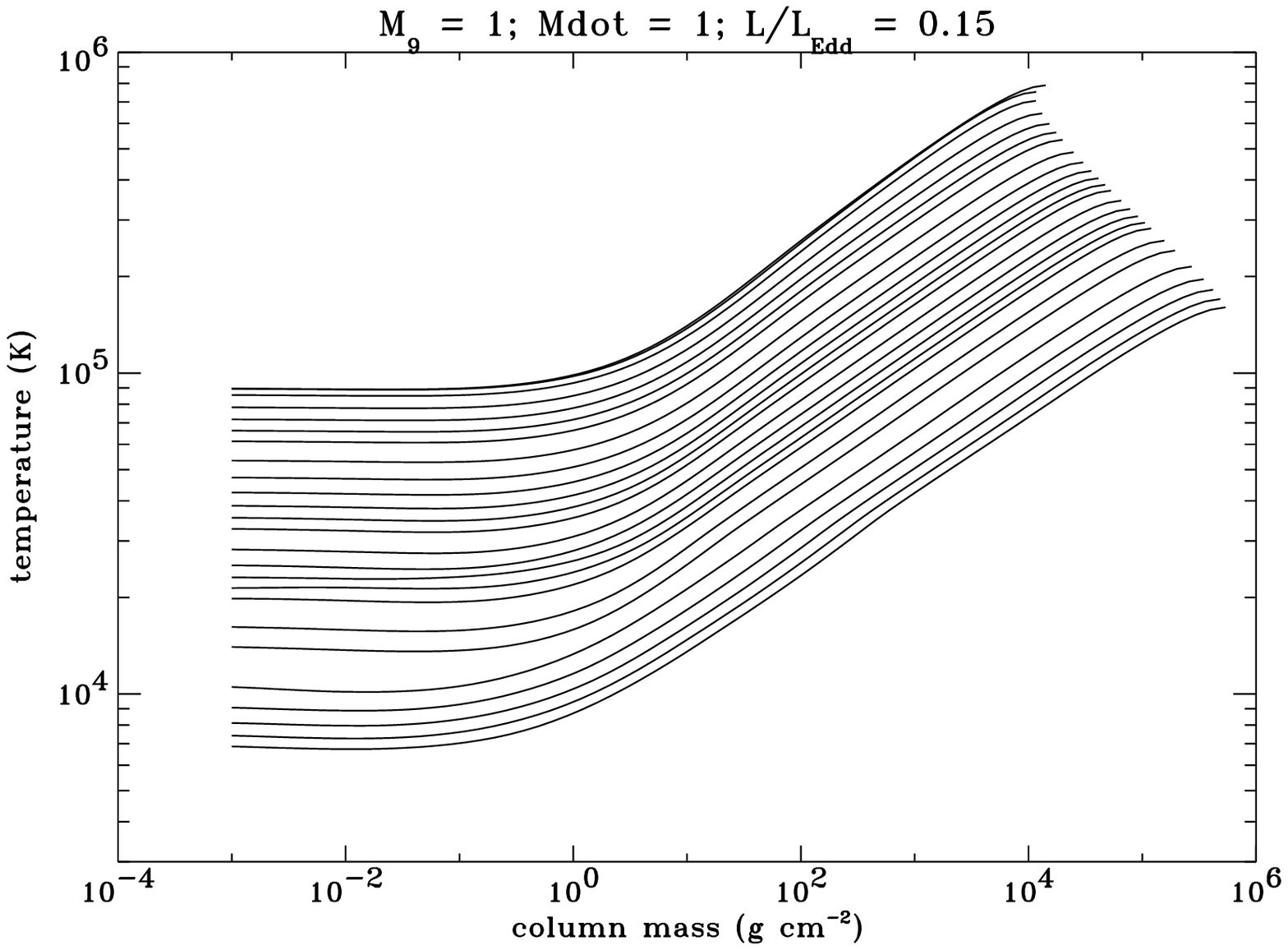,width=3.6in}} 
\noindent{
\scriptsize \addtolength{\baselineskip}{-3pt}
\vskip 1mm
\figcaption{
Temperature as a function of depth for the individual annuli. 
The curves correspond to radii $r/r_g$ (from top to bottom)
1.5, 2, 2.5, 3, 3.5, 4, 5, 6, 7, 8, 9, 10, 12, 14, 16, 18, 20, 25, 30,
40, 50, 60, 70, 80, and 90. \label{FIGTEMP}}
\vskip 3mm
\addtolength{\baselineskip}{3pt}
}

Figure \ref{FIGDENS} displays the run of mass density for the
individual annuli.  The central density is lowest for the inner annuli,
and increases with increasing distance from the black hole,
while the density close to the surface exhibits a more or less reverse 
behavior.  Note that the disk becomes optically thin below a column mass of
$m \simeq\chi_R^{-1}\simeq2.9$~g~cm$^{-2}$, where the density is substantially 
lower (in some cases by orders of magnitude) than the midplane density.  This
should be borne in mind when comparing our
results to those of previous workers who assumed constant density slabs
(e.g. Laor \& Netzer 1989).
An explanation of the behavior of the density is given in the
Appendix.

%
\vskip 2mm
\hbox{~}
\centerline{\psfig{file=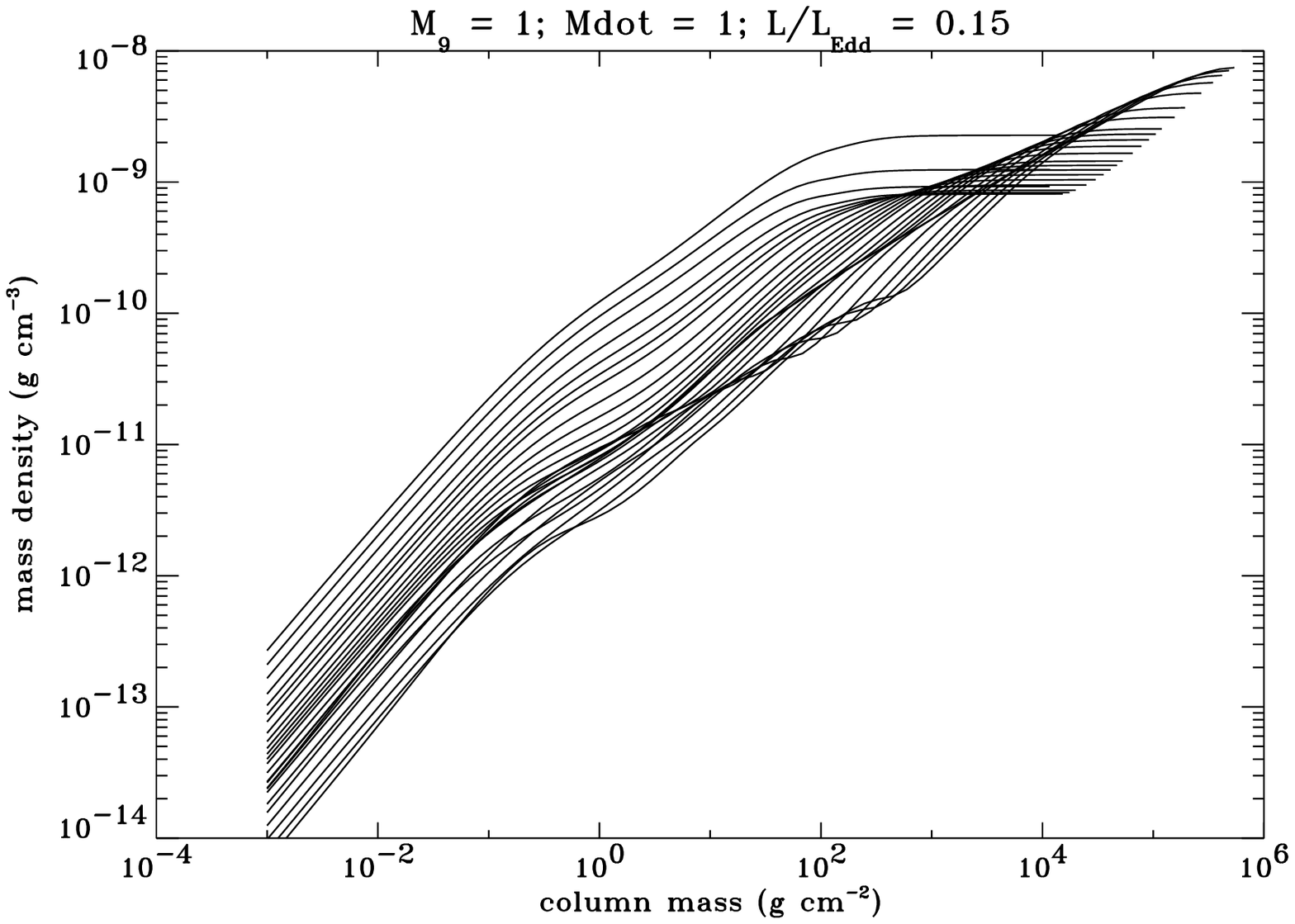,width=3.6in}} 
\noindent{
\scriptsize \addtolength{\baselineskip}{-3pt}
\vskip 1mm
\figcaption{Mass density as a function of depth for the individual annuli. 
The radii are the same as in Fig. \ref{FIGTEMP}. 
At the midplane, the highest density corresponds to the largest
$r/r_g$ (equal to 90), while the lowest density corresponds to
$r/r_g=2.5$. The two uppermost curves for $m < 10^3$ correspond to
$r/r_g=1.5$ and 2, respectively.
\label{FIGDENS}}
\vskip 3mm
\addtolength{\baselineskip}{3pt}
}

A more interesting behavior is exhibited by the H I and He II ground
state number densities, displayed in figures \ref{FIGNH1} and 
\ref{FIGNHE2}. For hot, inner annuli, hydrogen remains ionized throughout
the entire vertical extent of the disk, while starting with the
annulus at $r= 40  r_g$ (with $T_{\rm eff} \approx 15,000 $K), 
there is an appreciable
portion of neutral hydrogen in the outer layers ($m < 10^2$ g cm$^{-2}$).
Consequently, the Lyman edge opacity for these latter models becomes rather
large (see below).  Similarly, helium exhibits a transition, between
$r= 12\, r_g$ (with $T_{\rm eff} \simeq 36,000 $K) and  
$r= 14\, r_g$ (with $T_{\rm eff} \simeq 32,500 $K)
from being predominantly doubly
ionized, to a situation where the single ionized helium is a dominant
stage of ionization at some outer part of the disk. For the annuli
with $r< 40 r_g$ (with $T_{\rm eff} > 15,000$ K), helium remains singly
ionized throughout the entire upper part of the disk, while for more
distant, cooler annuli, neutral helium becomes more and more dominant
in the outer layers.
%
\vskip 2mm
\hbox{~}
\centerline{\psfig{file=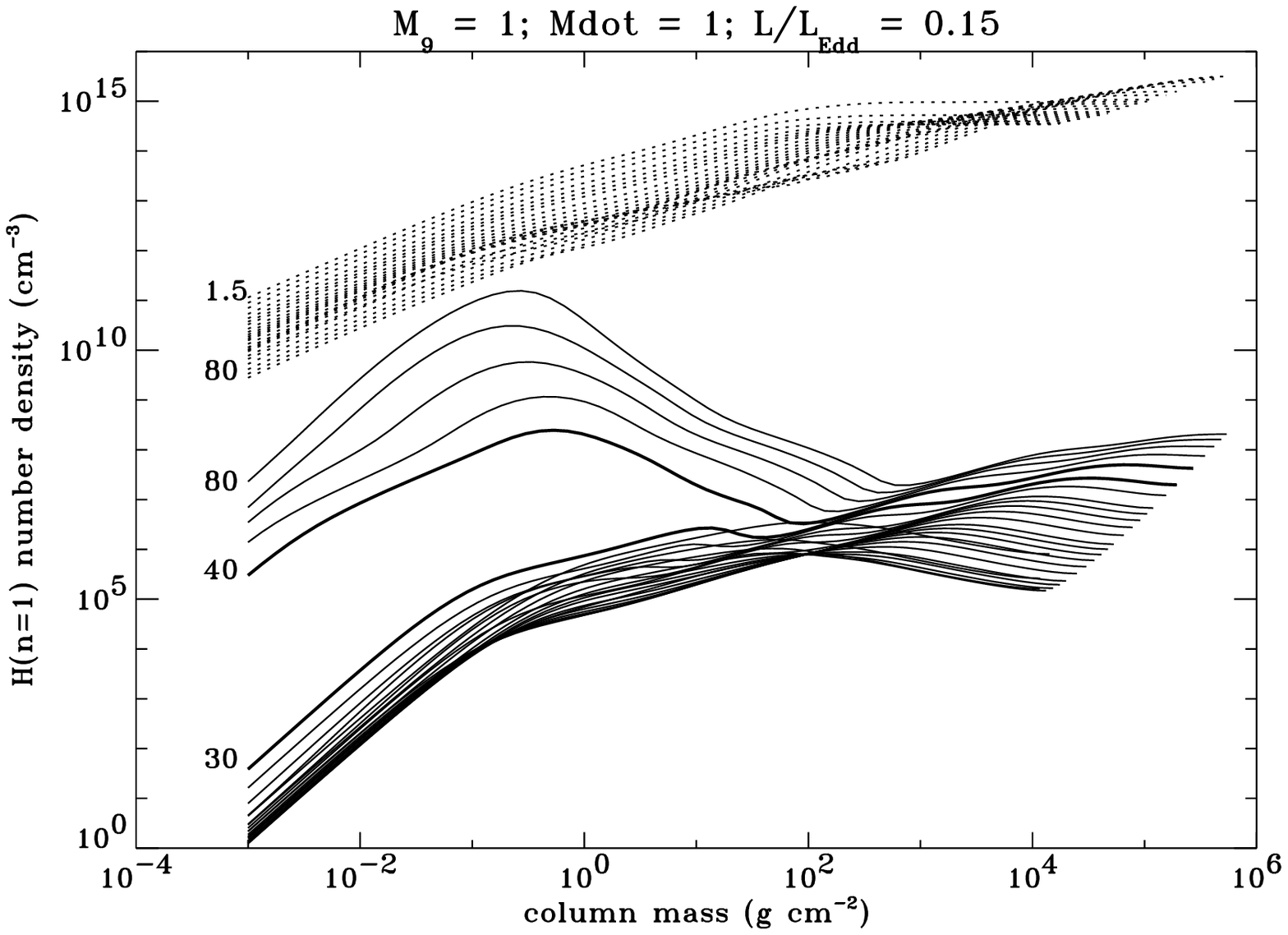,width=3.6in}} 
\noindent{
\scriptsize \addtolength{\baselineskip}{-3pt}
\vskip 1mm
\figcaption{Number density of the ground state of H I as a function
of depth (full lines), and the proton number density (dotted lines).
The radii are the same as in Fig. \ref{FIGTEMP}.
Selected models are labeled by the radial coordinate, $r/r_g$, and the
models for $r/r_g = 30$ and 40, which delimit the regions between the
full and partial ionization of hydrogen, are drawn by bold lines.
\label{FIGNH1}}
\vskip 3mm
\addtolength{\baselineskip}{3pt}
}
%
%
\vskip 2mm
\hbox{~}
\centerline{\psfig{file=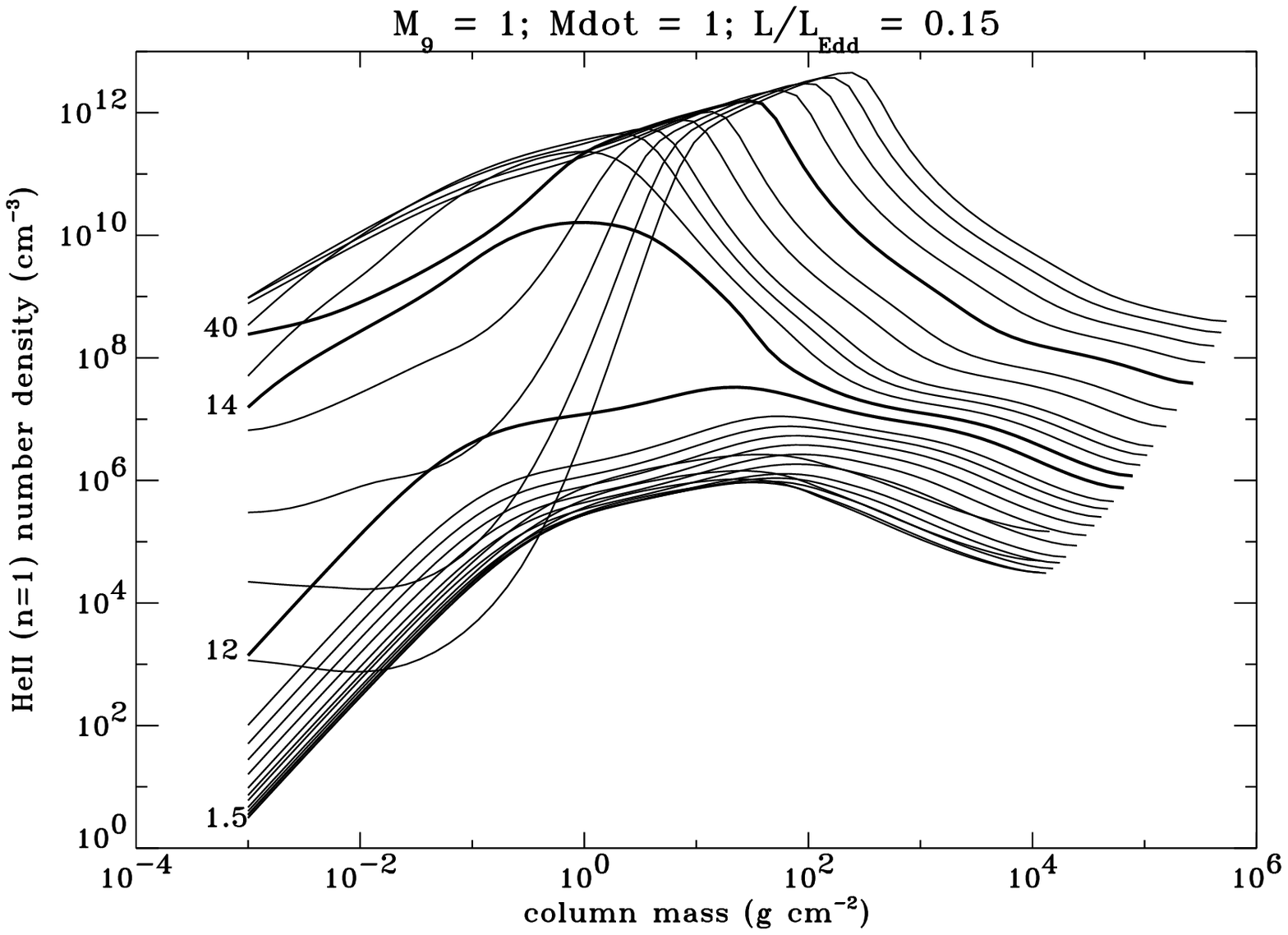,width=3.6in}} 
\noindent{
\scriptsize \addtolength{\baselineskip}{-3pt}
\vskip 1mm
\figcaption{Number density of the ground state of He II as a function
of depth. The radii are the same as in Fig. \ref{FIGTEMP}.
Selected models are labeled by the radial coordinate, $r/r_g$, and the
models for $r/r_g = 12$, 14, and 40, are drawn by bold lines.
\label{FIGNHE2}}
\vskip 3mm
\addtolength{\baselineskip}{3pt}
}
Figure \ref{FIGFLX1} displays the emergent flux for all annuli. The upper
panel shows the non-LTE models, while the lower panel shows the LTE models.
The behavior of the emergent flux is analogous to that discussed at length
in Papers I and II. Non-LTE models exhibit the He II Lyman edge in
emission for $r< 12 \, r_g$ (with $T_{\rm eff} > 36,000$K), while for more
distant annuli the flux in the He II Lyman continuum drops to almost
zero. This is a consequence of the transition from the dominance of 
double-ionized helium to singly-ionized helium, as displayed in
figure \ref{FIGNHE2}. Analogously, the neutral-hydrogen annuli for
$r> 40\, r_g$, with $T_{\rm eff} > 15,000$ K, (see figure \ref{FIGNH1}) show 
a significant hydrogen Lyman edge in absorption. 

A comparison between LTE and non-LTE results reveals several interesting 
effects: in LTE, the He II Lyman edge appears to be in weaker
emission or even in absorption for
hot, doubly-ionized, annuli. Most importantly, non-LTE effects reduce
the hydrogen Lyman edge (cf. Sun \& Malkan 1989, Shields \& Coleman 1994,
St\"orzer et al. 1994, Papers I and II):  the edge predicted in non-LTE
models is typically in weaker emission for emission edges, and in weaker 
absorption for absorption edges.  We discuss the behavior of the hydrogen
Lyman edge in the full disk-integrated spectra in more detail in Sect. 3.5
below.

Finally, figure \ref{FIGFLX2} shows a comparison of predicted local non-LTE
flux and the blackbody flux corresponding to the same effective
temperature.  As already discussed in Paper I, the non-LTE models
exhibit a much wider spectral energy distribution than blackbodies.
Consequently, an often-used mapping of a given spectral region to a certain
radial position in the disk, which may be used for a blackbody flux
because of its sharp variation with  frequency, is much less satisfactory
for more realistic non-LTE vertical structure models. Another crucial
feature of the present models is that compared to blackbodies the
energy is redistributed.  At low temperatures, absorptive opacity is
relatively important, so that flux is shifted from frequen-
%
\hbox{~}
\centerline{\psfig{file=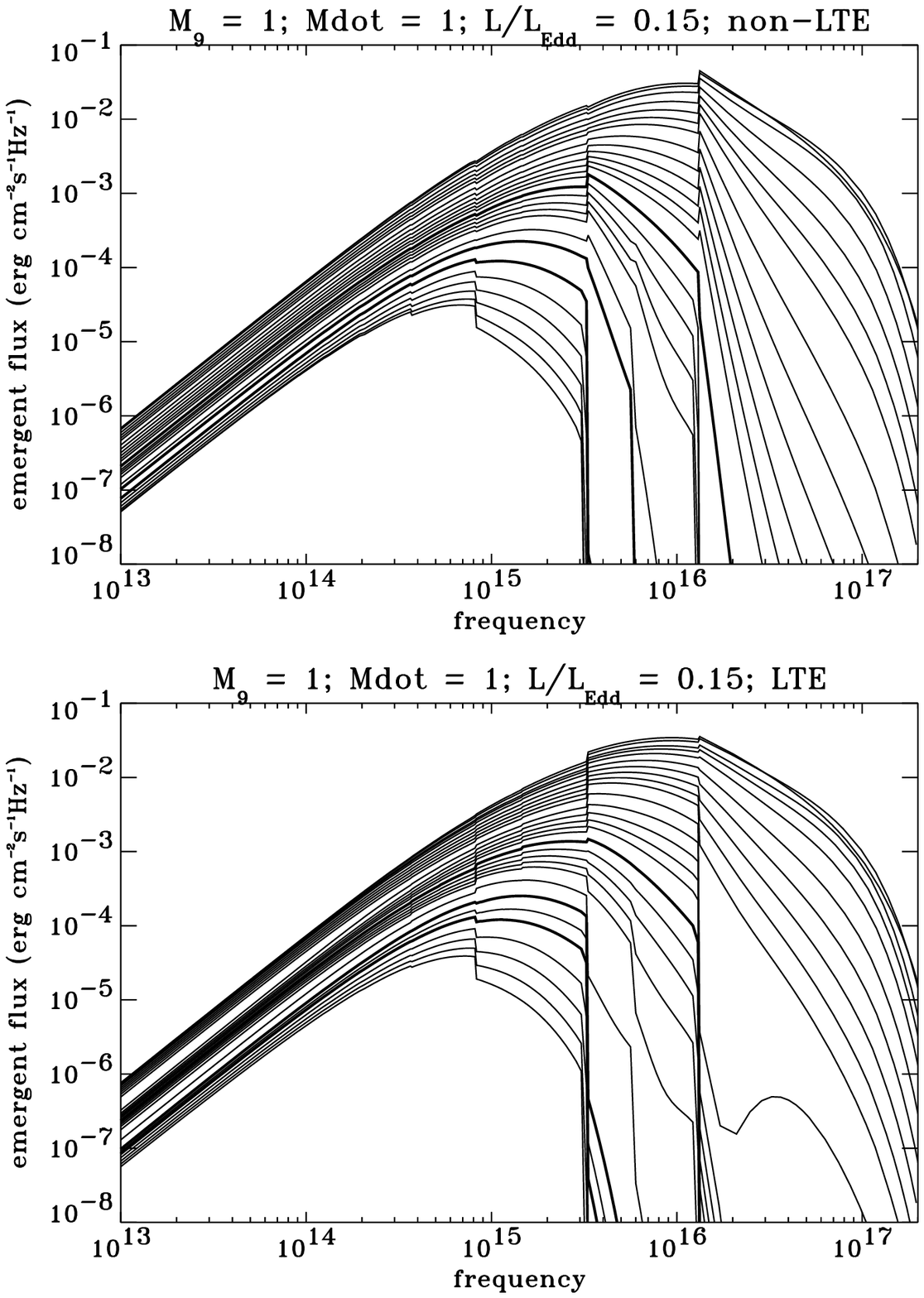,width=3.6in}} 
\noindent{
\scriptsize \addtolength{\baselineskip}{-3pt}
\vskip 1mm
\figcaption{Predicted local emergent flux (in erg cm s$^{-1}$ Hz$^{-1}$)
for the individual annuli for non-LTE models (upper panel), and LTE models 
(lower panel).  The models for $r/r_g = 12$, 30, and 40, are drawn by bold 
lines. \label{FIGFLX1}}
\vskip 3mm
\addtolength{\baselineskip}{3pt}
}
%
%
\vskip 2mm
\hbox{~}
\centerline{\psfig{file=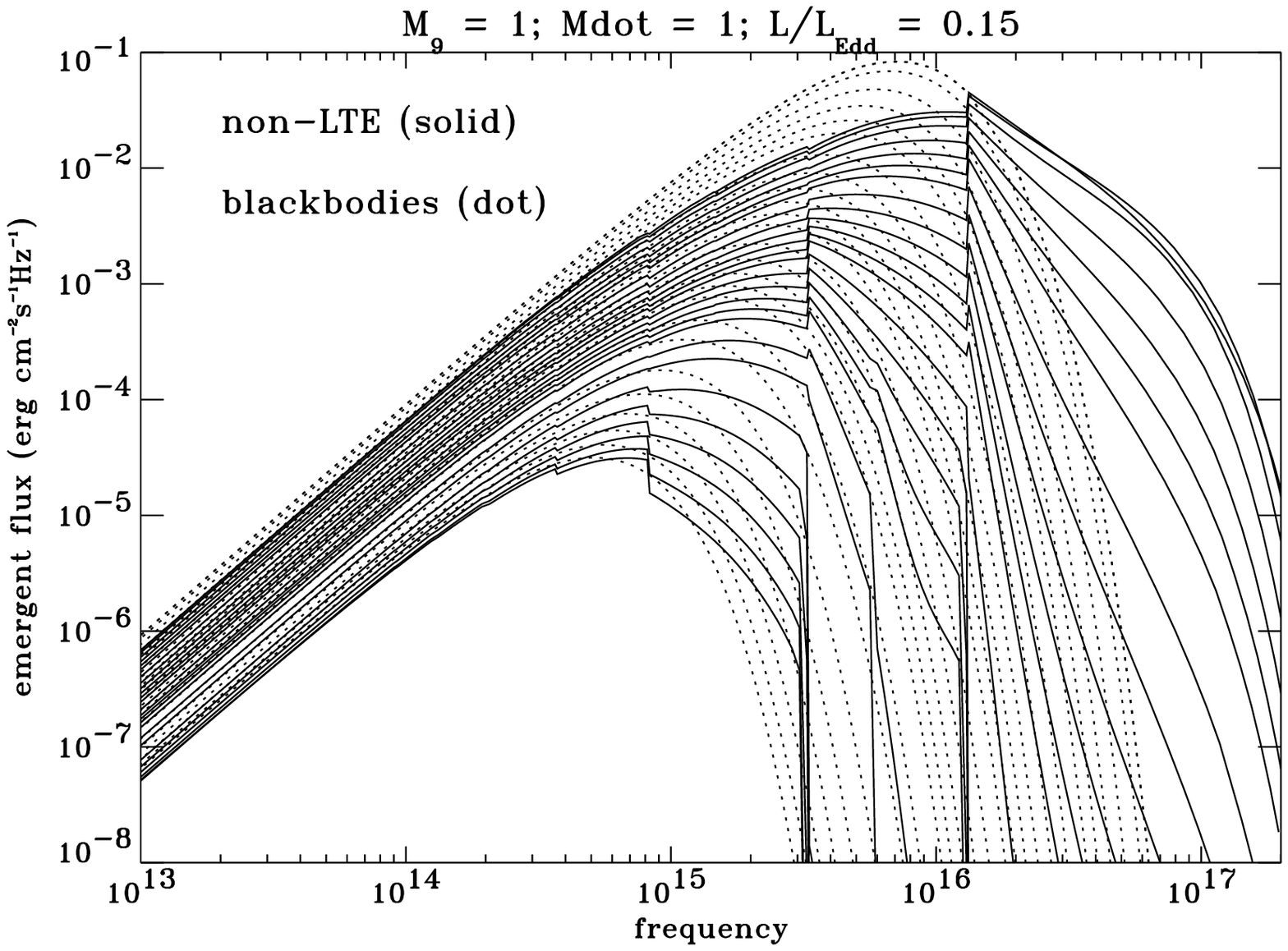,width=3.6in}} 
\noindent{
\scriptsize \addtolength{\baselineskip}{-3pt}
\vskip 1mm
\figcaption{
Predicted local emergent flux (in erg cm s$^{-1}$ Hz$^{-1}$)
for the individual annuli for non-LTE models (full lines), together with a 
blackbody flux corresponding to the same effective temperature as the
given annulus (dotted lines).
\label{FIGFLX2}
}
\vskip 3mm
\addtolength{\baselineskip}{3pt}
}
\noindent cies where the opacity is high to frequencies where it is low; i.e., away from ionization
edges.  Another way of viewing this effect is to note that in bands
where the opacity is low, escaping photons are created deeper inside the
disk where the temperature is higher.  At high temperatures, the opacity
becomes scattering dominated.  In these rings, there is a trade-off
between the effect just mentioned (lower opacity bands allow one to see
photons created at higher temperature) and scattering blanketing (which
retards the escape of photons created deep inside).  The latter effect
is the one that creates a ``modified blackbody" spectrum when there is
no temperature gradient.  For much of the parameter range of interest,
the net effect is to shift flux toward higher frequencies.

\subsection{Consistency of Models}

Having computed the detailed vertical structure of the annuli, we have
to address the question of self-consistency of the model assumptions.

First, we check that the computed disks are indeed geometrically thin,
i.e., the disk height, $H$, is much smaller than the radial coordinate $r$.
In figure \ref{FIGZ1} we show
the ratio of the disk height to the radial coordinate, $H/r$, as a function
of the radial coordinate (in units of gravitational radius), for a disk
with $M_9 = 1$, and for various values of $\dot M$ (or luminosity).
The behavior of disks for other values of the black hole mass is almost
identical.  (This is expected for radiation pressure supported, electron
scattering dominated disks, because then $H/r$ can be written as
$L/L_{\rm Edd}$ times a function of $r/r_g$, cf. equation 53 of Paper II.)
Only for the most luminous disk does the ratio $H/r$ approach 10\% at 
$r/r_g= 2.5$.  The height of other annuli is lower, and the maximum height
for less luminous disks is progressively lower.  These results are in good
agreement with the older models of Laor \& Netzer (1989, cf. their figure 1
and the surrounding discussion).  We therefore conclude that our disks do
indeed satisfy the thin disk approximation.

%
\vskip 2mm
\hbox{~}
\centerline{\psfig{file=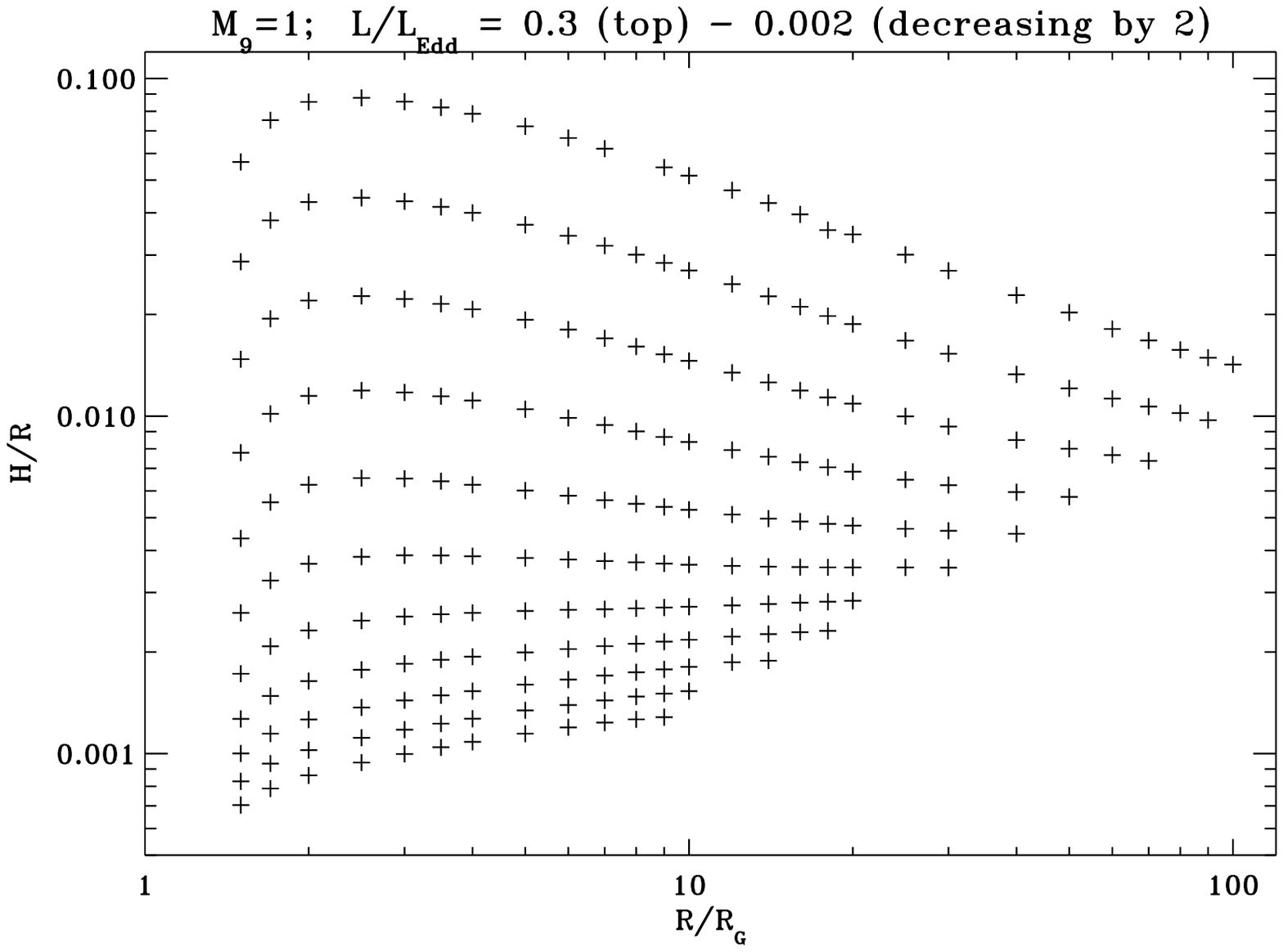,width=3.6in}} 
\noindent{
\scriptsize \addtolength{\baselineskip}{-3pt}
\vskip 1mm
\figcaption{
The ratio of the disk height to the radial coordinate, $H/r$, as a 
function
of the radial coordinate (in units of gravitational radius), for a disk
with $M_9 = 1$, and for various values of $\dot M$ (or luminosity).
The crosses represent the actual computed models of the individual annuli.
\label{FIGZ1}
}
\vskip 3mm
\addtolength{\baselineskip}{3pt}
}

Another important concern is the presence of vertical density inversions
within the disk.  Since we neglect convection, sharp temperature
gradients and density inversions are found at ionization transitions occurring
in regions where gas pressure contributes significantly to support against
gravity.  In figure \ref{FIGCOOL} we display the structure of several annuli that
illustrate this behavior. The hottest annulus shown in that figure
($r/r_g= 50$) has $T_{\rm eff} = 9400$ K; it is the outermost
ring with no density inversion.  Cooler, more distant annuli show a
progressively stronger inversion.  The inversion is created by the abrupt fall
in electron density when H recombines; in order to maintain the pressure
gradient required for hydrostatic balance, the mass density must increase to
compensate. 
A sharp rise of local temperature, displayed in the upper panel of figure
\ref{FIGCOOL}, is caused by a rapid increase of opacity when going inward
due to an ionization front.
The temperature is essentially a function of the Rosseland mean opacity,
$T \propto \tau_{\rm Ross}^{1/4}$, but since $\tau_{\rm Ross}$ is a sharp
function of the column mass $m$, the function $T(m)$ also exhibits a sharp
increase with $m$.

%
\vskip 2mm
\hbox{~}
\centerline{\psfig{file=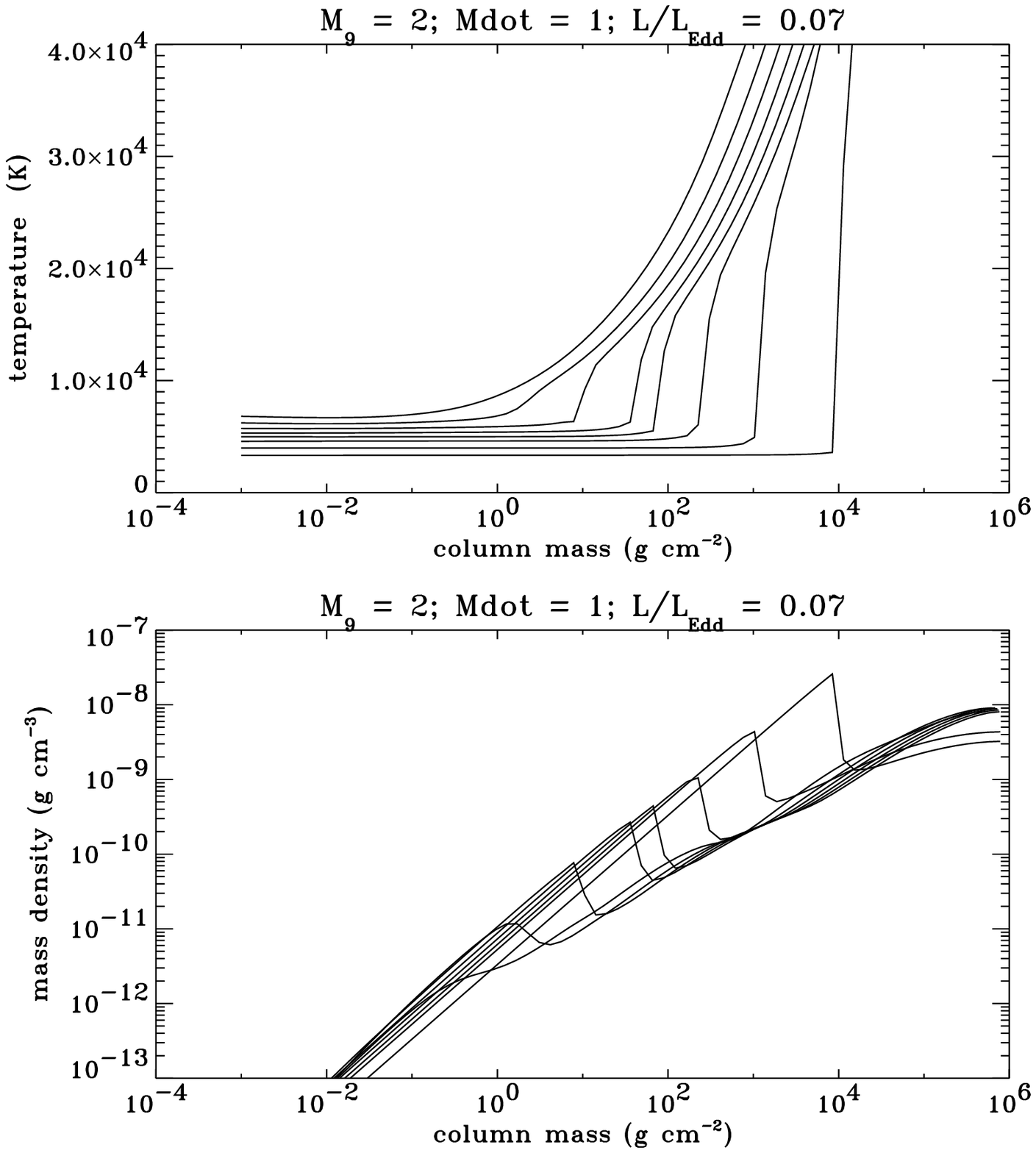,width=3.6in}} 
\noindent{
\scriptsize \addtolength{\baselineskip}{-3pt}
\vskip 1mm
\figcaption{
Upper panel: temperature as a function of position 
for a disk with $M_9=2$ and $\dot M = 1 M_\odot$ yr$^{-1}$,
for ``cool'' annuli (for radii 50, 60, 70, 80, 90, 100, 120, and 160 $r_g$.
The corresponding $T_{\rm eff}$'s are 9400, 8300, 7400, 6700, 6200, 5700,
5000, and 4100 K.
The lower panel: mass density for the same annuli.
\label{FIGCOOL}
}
\vskip 3mm
\addtolength{\baselineskip}{3pt}
}

These models should be taken with caution.
In the absence of detailed hydrodynamical calculations which would
allow for convective instability and determine the vertical structure
properly, we do not know what is the correct emergent radiation. 
We can nevertheless obtain a rough indication of model uncertainties by 
comparing the predicted flux from the non-LTE models that neglect convection, 
and a blackbody flux corresponding to the same effective temperature.

We present in figure \ref{FLUXCOOL} emergent flux for three selected annuli
shown in figure \ref{FIGCOOL}, for $r/r_g = 50, 90, 160$, with corresponding
effective temperature equal to 9400, 6200, 4100 K, respectively. 
The predicted flux for non-LTE models is shown together with the
corresponding blackbody flux.  We see that the models at the hotter end
of the density inversion sequence exhibit an appreciable Balmer edge,
while the cooler models are reasonably well approximated by blackbodies. 
We thus feel that approximating even cooler models, which we do not compute
here, by blackbodies is probably not worse than other approximations that 
underlie the entire calculation.
%
\vskip 2mm
\hbox{~}
\centerline{\psfig{file=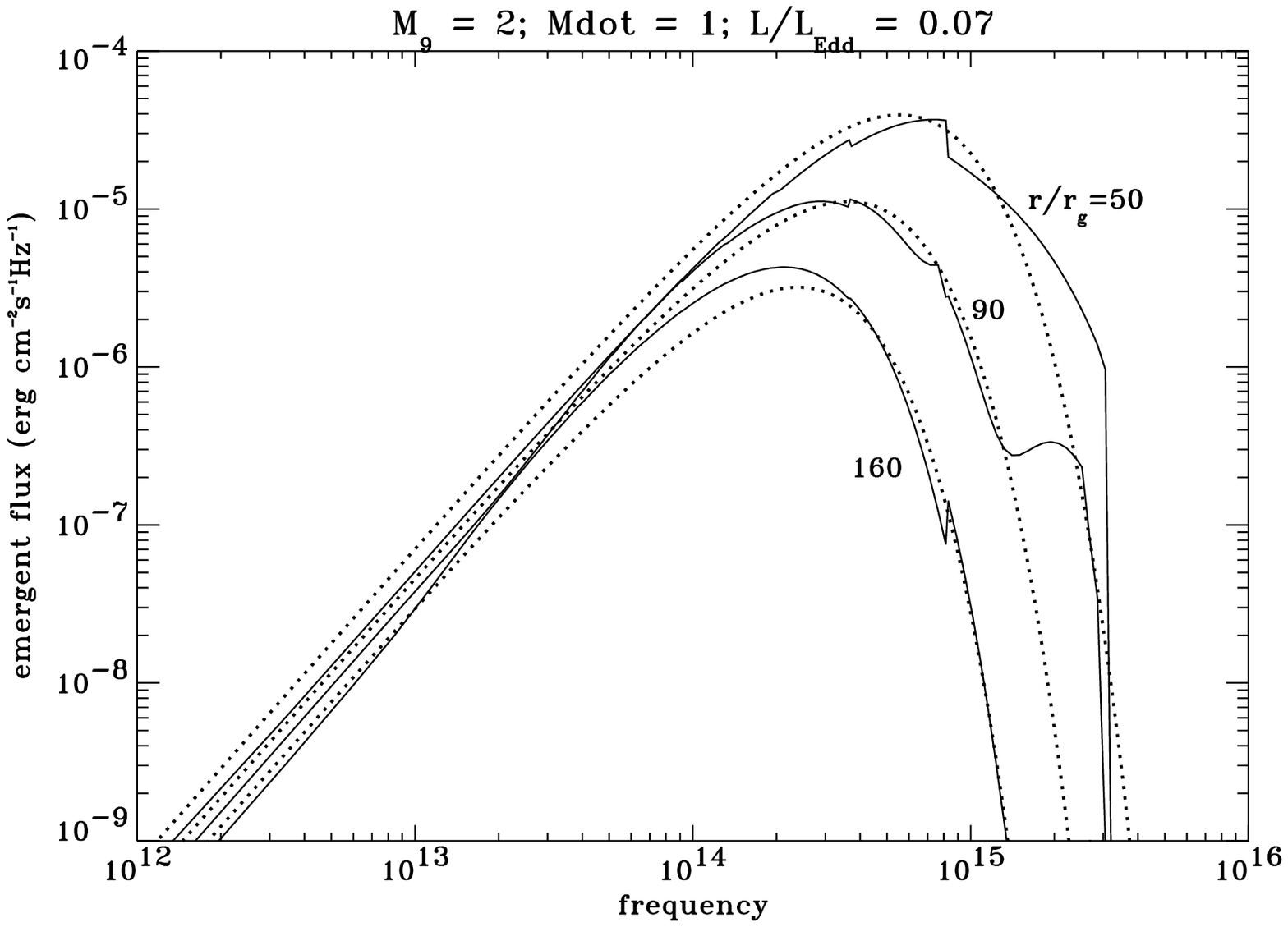,width=3.6in}} 
\noindent{
\scriptsize \addtolength{\baselineskip}{-3pt}
\vskip 1mm
\figcaption{
Predicted local emergent flux (in erg cm s$^{-1}$ Hz$^{-1}$)
for three selected annuli shown in Fig. \ref{FIGCOOL}. 
Full lines:  non-LTE models; dotted lines: 
blackbody flux corresponding to the same effective temperature as the
given annulus. The curves are labeled by the value of $r/r_g$ = 50, 90, 160;
the corresponding $T_{\rm eff} = 9400, 6200, 4100$ K, respectively.
\label{FLUXCOOL}
}
\vskip 3mm
\addtolength{\baselineskip}{3pt}
}

\subsection{Disk-integrated Spectra}

In this section, we present disk-integrated spectra for selected disk
models. The full grid of models is not presented here, but is available
to interested researchers upon request.

Note that in all the spectral energy distributions shown in this paper,
the quantity $L_\nu$ is the specific luminosity that an observer along
a particular viewing angle would infer the source to have if it were
isotropic, i.e. if $F_\nu$ is the measured specific flux and $d$ is
the distance to the source, then $L_\nu\equiv4\pi d^2F_\nu$.
In the subsequent plots, we display,
as is customary, the quantity $\nu L_\nu$, which represents a luminosity per
unit logarithmic interval of frequency (photon energy).

As discussed above, we consider the spectra of the outer annuli which
were not computed  (for $T_{\rm eff} < 4000$ K) to be given
by the black-body energy distribution corresponding to the effective
temperature of the annulus.
An important parameter is the outer cutoff of
the disk. In order to avoid problems with an improper choice of the
outer cutoff, we have chosen the cutoff radius $r_{\rm out}$ in such a
way that $T_{\rm eff}(r_{\rm out})$ is equal to a specific limiting
temperature, $T_{\rm lim}$.  We have chosen $T_{\rm lim} = 1000$ K,
which guaranties that the total emergent flux at $\nu = 10^{14}$, which
is the lowest frequency considered in our integrated disk spectra, 
is not significantly influenced by cooler annuli with 
$T_{\rm eff} < T_{\rm lim}$ to within a few per cent.
%
\vskip 2mm
\hbox{~}
\centerline{\psfig{file=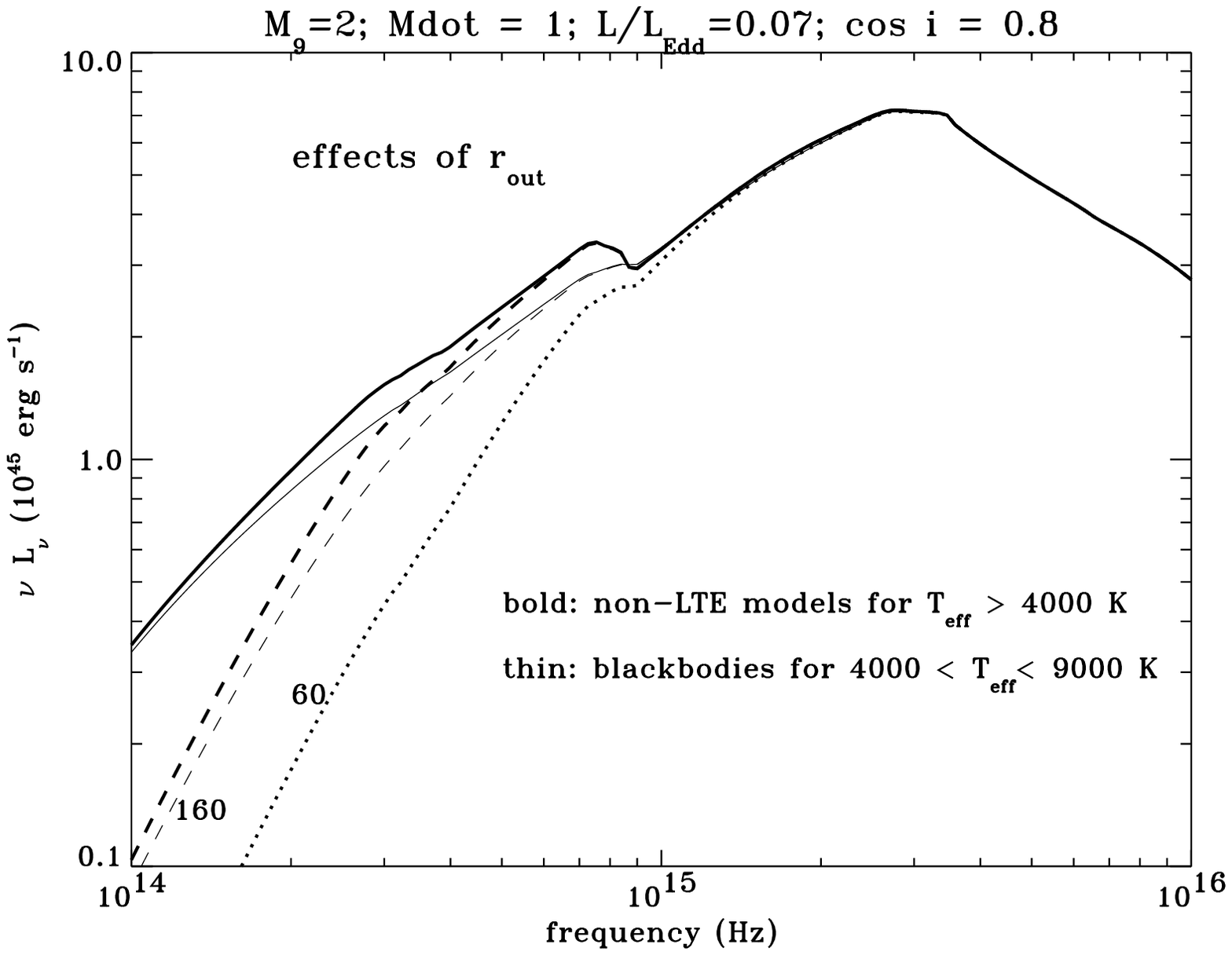,width=3.6in}} 
\noindent{
\scriptsize \addtolength{\baselineskip}{-3pt}
\vskip 1mm
\figcaption{
Integrated spectral energy distribution $\nu L_\nu$
for a disk with $M_9 =2$ and
$\dot M = 1 \, M_\odot$ yr$^{-1}$ ($L/L_{\rm Edd} = 0.07$), with inclination
$\cos i= 0.8$, and with
various choices of $r_{\rm out}$.
Solid lines: the automatic choice of $r_{\rm out}$, corresponding to 
$T_{\rm lim} = 1000$ K (in this case, $r_{\rm out}/r_g\approx 1060$).
Dashed lines:  $r_{\rm out}/r_g=160$;
Dotted line: $r_{\rm out}/r_g=60$; the curves are labeled by the value of
$r_{\rm out}/r_g$).
Bold lines correspond to spectra computed for non-LTE
models down to $T_{\rm eff} = 4000$ K; 
thin lines show the spectra computed by replacing the local spectra of 
annuli with $T_{\rm eff} < 9000$ K by blackbodies.
\label{FIGC0}
}
\vskip 3mm
\addtolength{\baselineskip}{3pt}
}
Figure \ref{FIGC0} shows the effect of the outer cutoff, as well as of the
degree of approximation in the treatment of cool annuli, for the disk
model displayed in figure \ref{FIGCOOL} ($M_9 =2$; 
$\dot M = 1 \, M_\odot$ yr$^{-1}$, i.e., $L/L_{\rm Edd} = 0.07$). Three cutoff
radii are shown: $r_{\rm out}/r_g = 60$, which is the radius where the
density inversion sets in; $r_{\rm out}/r_g = 160$ (the radius where
$T_{\rm eff}$ reaches the our minimum value of 4000 K), and finally
the default value which corresponds to $T_{\rm lim} = 1000$ K, which in this
case happens at $r_{\rm out}/r_g \approx 1060$.
Neglecting all annuli cooler than 9000 K (dotted line) does not influence the 
flux blueward of the Balmer limit, but the optical and IR flux are seriously 
underpredicted.
Neglecting all non-computed annuli cooler than 4000 K (dashed lines) produces
the correct flux in the region of the Balmer edge, and very nearly the correct
flux in the optical range ($\nu > 4 \times 10^{14}$ Hz).  The effect
of uncertainties in the models with density inversion ($T_{\rm eff} < 9000$ K)
is estimated by comparing bold lines, which show spectra computed for non-LTE
models down to $T_{\rm eff} = 4000$ K, with the corresponding thin lines,
showing the predicted integrated spectra when
all annuli with $T_{\rm eff} \leq 9000$ K are assumed to emit locally as
blackbodies.  The maximum difference in the integrated spectra is found
redward of the Balmer edge; the non-LTE models of cool annuli produce an
integrated flux that is about 16 \% higher than that corresponding
to replacing the $T_{\rm eff} < 9000$ K annuli by blackbodies. 
Therefore, the effect is not very large, which gives us confidence that
our overall integrated spectra are not significantly influenced by the
uncertainties associated with the density-inversions present in the cool
annuli.  However, the predicted feature at the 
Balmer edge should be viewed with caution.

We first present integrated spectra for a disk of given mass and luminosity, 
$M_9 = 1$, $L/L_{\rm Edd} = 0.15$ (i.e. $\dot M =1\, M_\odot$ yr$^{-1}$), and
for various values of the inclination angle $i$, ranging from $\cos i = 0.99$
(i.e., the disk seen almost face-on), to $\cos i = 0.01$ (the disk seen almost
edge-on) -- figure \ref{FIGTOT1}. We can clearly see the impact of relativistic
boosting and aberration: the flux at the highest frequencies (radiated where the
disk is most relativistic) is boosted strongly for viewing angles near
the disk plane.  Also for such viewing angles, gravitational
light bending counteracts the Newtonian ``$\cos\theta$" projection
effect.
%
\hbox{~}
\centerline{\psfig{file=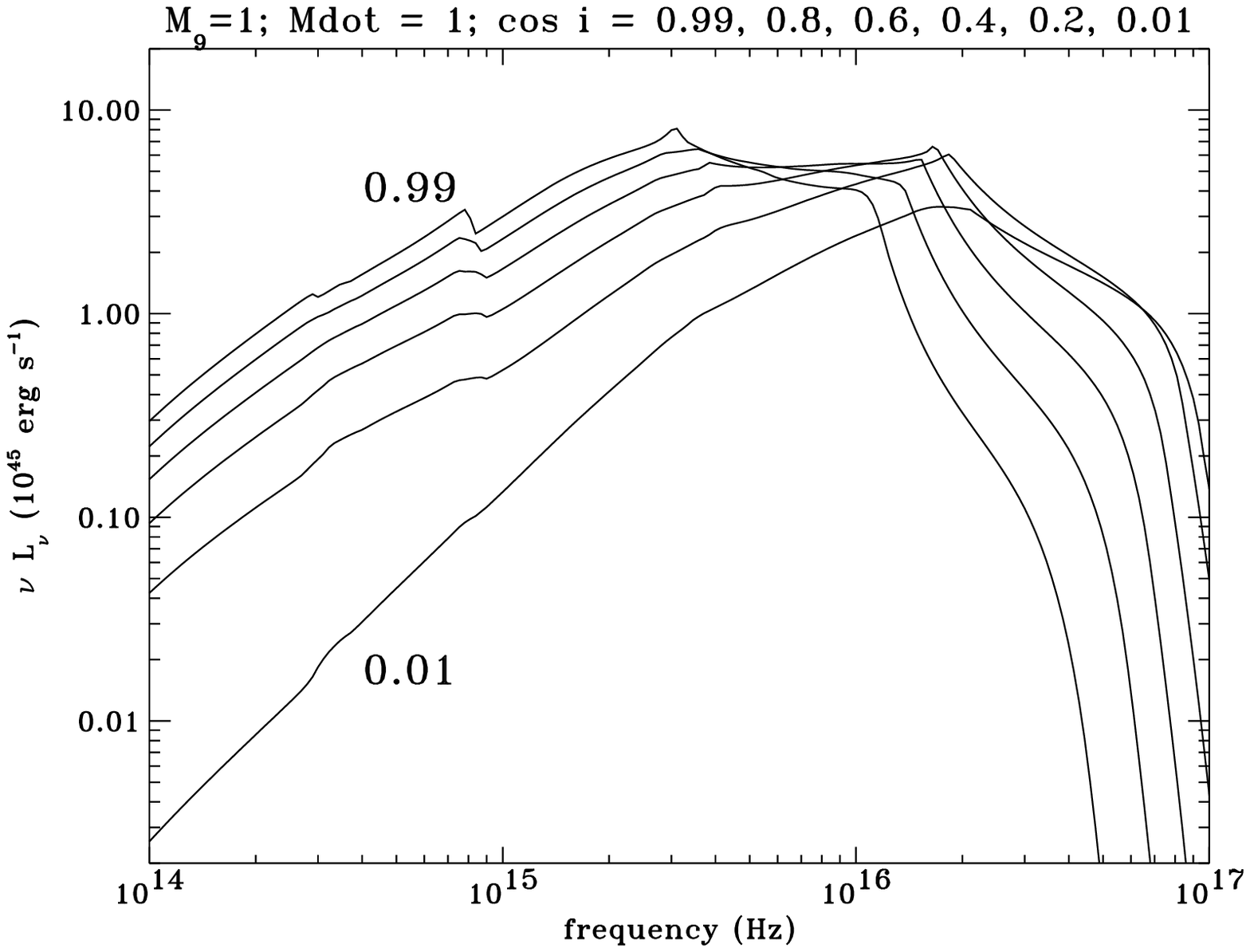,width=3.6in}} 
\noindent{
\scriptsize \addtolength{\baselineskip}{-3pt}
\vskip 1mm
\figcaption{
Integrated spectral energy distribution $\nu L_\nu$ 
(in erg s$^{-1}$) for a disk with $M_9 = 1$, $\dot M =1 M_\odot$ yr$^{-1}$,
and for various inclinations.
\label{FIGTOT1}
}
\addtolength{\baselineskip}{3pt}
}
%

%
\hbox{~}
\centerline{\psfig{file=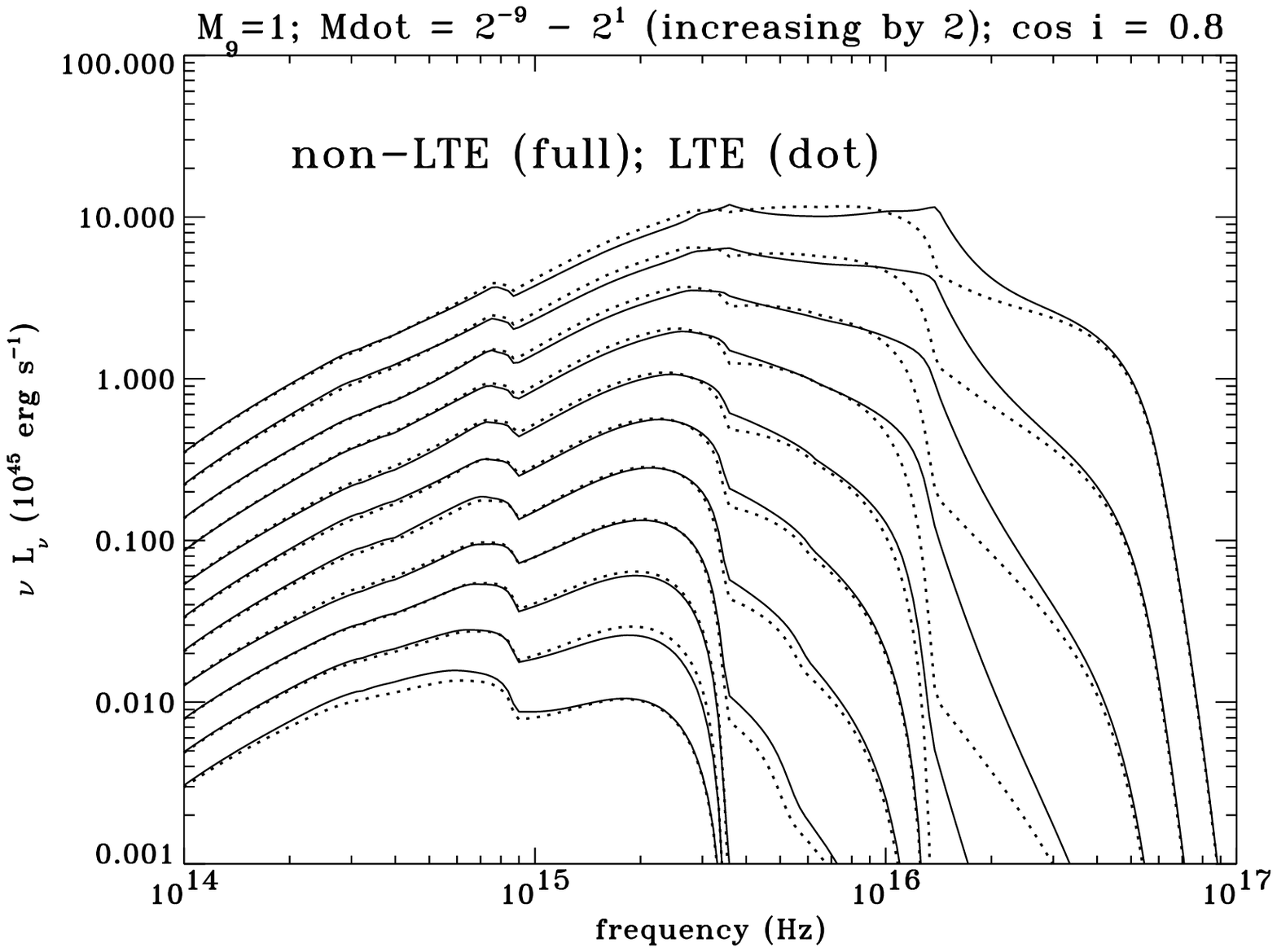,width=3.6in}} 
\noindent{
\scriptsize \addtolength{\baselineskip}{-3pt}
\vskip 1mm
\figcaption{
Integrated spectral energy distribution $\nu L_\nu$ 
(in erg s$^{-1}$) for a disk with $M_9 = 1$, $\cos i = 0.8$,
and for various values of the mass accretion rate (luminosity),
ranging from $L/L_{\rm Edd} = 0.3$ (i.e., $\dot M = 2 M_\odot$ yr$^{-1}$)
to $L/L_{\rm Edd} = 3 \times 10^{-4}$ 
(i.e., $\dot M = 2^{-9} M_\odot$ yr$^{-1}$).
Full lines: non-LTE models; dotted lines: LTE models.
\label{FIGTOT2}
}
\addtolength{\baselineskip}{3pt}
}
In the following, we present the integrated spectra for one value of
inclination.  We chose $\cos i = 0.8$ (i.e., $i\approx 37^\circ$), which is 
a value relatively close to face-on, and which thus may serve as a typical
value for type 1 AGN and QSO's based on unification arguments (e.g., Krolik
1999a).
Figure \ref{FIGTOT2} shows the integrated spectra for one particular value of
the black hole mass, $M_9 = 1$, 
and for various luminosities (mass accretion rates).
Full lines display non-LTE models, while the dotted lines display LTE
model predictions. The spectral energy distribution is hardest for
the largest luminosity. The non-LTE effects are most important
in the He II Lyman continuum ($\nu > 1.36 \times 10^{16}$ Hz), and also
for predicting the detailed shape of the hydrogen Lyman edge for intermediate
and low luminosities.

%
\vskip 2mm
\hbox{~}
\centerline{\psfig{file=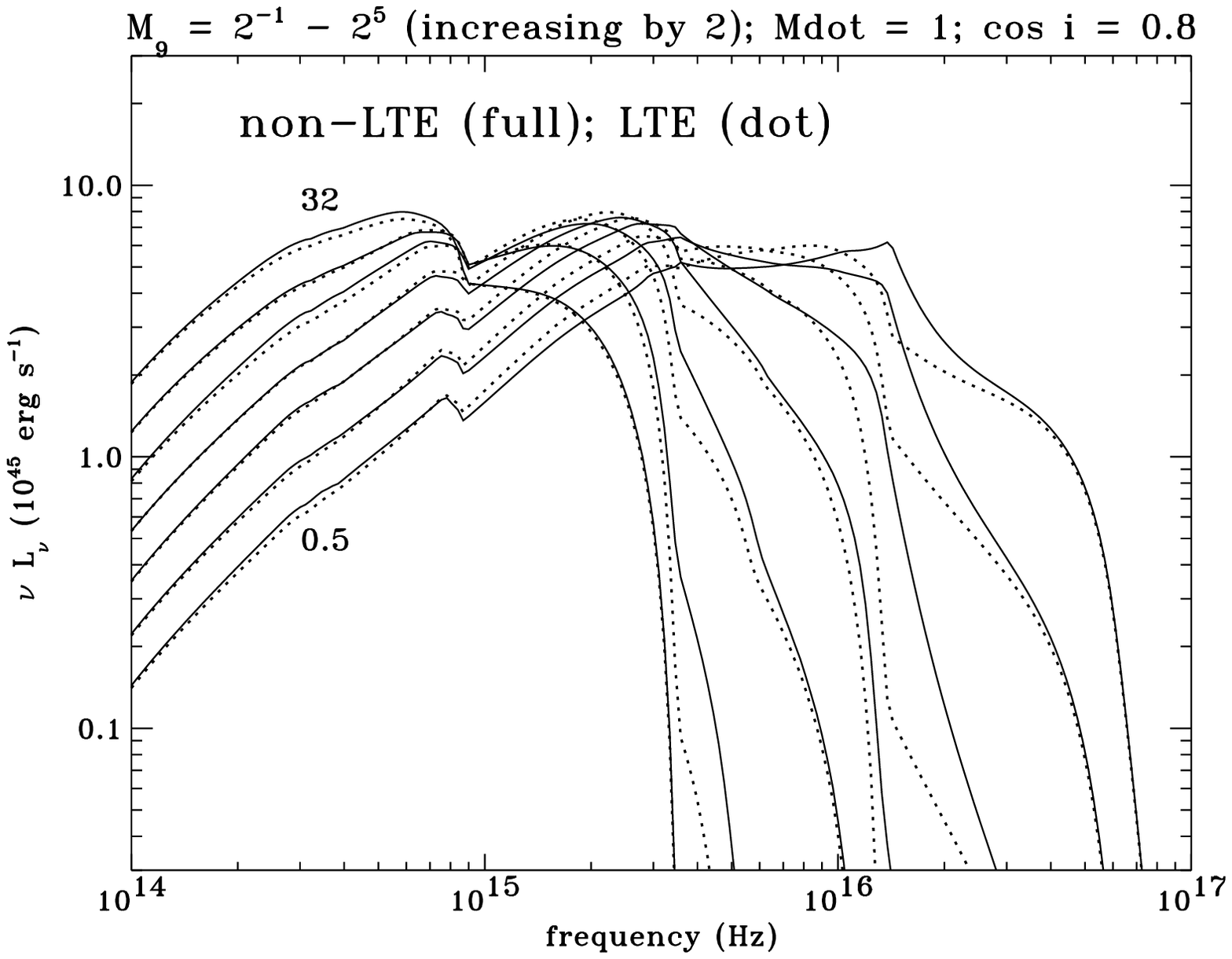,width=3.6in}} 
\noindent{
\scriptsize \addtolength{\baselineskip}{-3pt}
\vskip 1mm
\figcaption{
Integrated spectral energy distribution $\nu L_\nu$ 
(in erg s$^{-1}$) for a disk with $\dot M = 1 M_\odot$ yr$^{-1}$,
$\cos i = 0.8$, and for various values of the black hole mass, $M_9$,
ranging from $M_9 = 32$ to 0.5.
\label{FIGTOT3}
}
\vskip 3mm
\addtolength{\baselineskip}{3pt}
}
In figure \ref{FIGTOT3} we display the sequence of predicted spectra for
models with a fixed mass accretion rate (i.e., total luminosity)
and varying black hole mass. The energy distribution is
progressively shifted towards more energetic photons for lower masses,
because disks around less massive holes have smaller radiating surface areas.
The non-LTE effects are important for all
disks; for hotter ones the largest departures from LTE are seen
in the He II Lyman continuum, and in cooler ones in the hydrogen
Lyman continuum.

%
\vskip 2mm
\hbox{~}
\centerline{\psfig{file=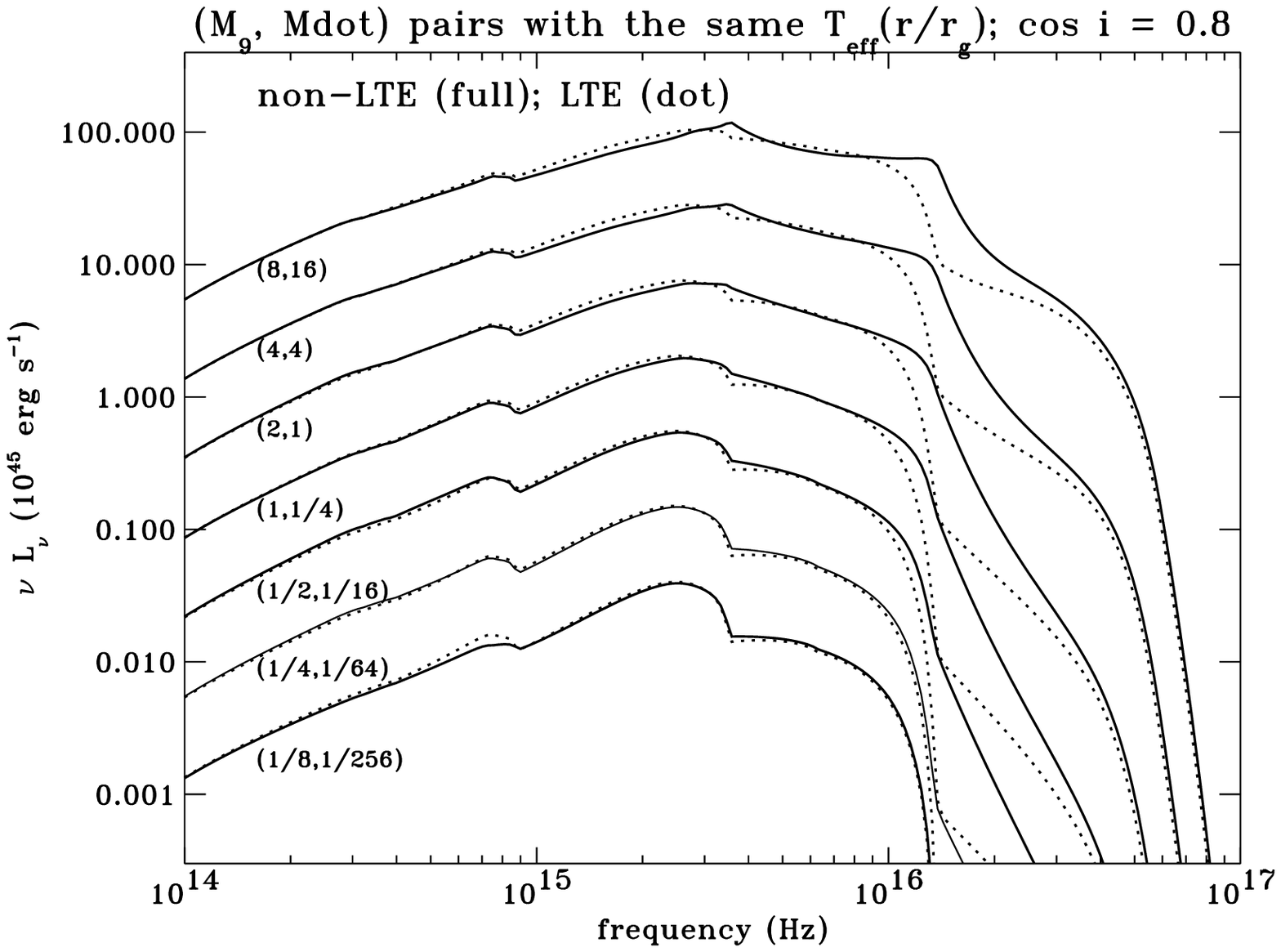,width=3.6in}} 
\noindent{
\scriptsize \addtolength{\baselineskip}{-3pt}
\vskip 1mm
\figcaption{
Integrated spectral energy distribution $\nu L_\nu$ 
(in erg s$^{-1}$) for the pairs of $(M_9, \, \dot M)$ which have the
same effective temperature distribution, $T_{\rm eff}(r/r_g)$, and
which should therefore have a rather similar spectrum shape.
The curves are labeled by the values of $(M_9, \dot M)$.
Full lines: non-LTE models; dotted lines: LTE models.
\label{FIGTOT4}
}
\vskip 3mm
\addtolength{\baselineskip}{3pt}
}
Figure \ref{FIGTOT4} shows the spectral energy distribution for a sequence
of disk models which have the same $T_{\rm eff}(r/r_g)$ distribution.
Since
\begin{equation}
T_{\rm eff}^4 \propto M \dot M r^{-3} R_R(r/r_g) \propto
M^{-2} \dot M (r/r_g)^{-3} R_R(r/r_g)\, ,
\end{equation}
the same $T_{\rm eff}$ distribution is 
obtained for models with fixed $\dot M/M^2$.  If disks
radiated as blackbodies, all the spectra of the sequence would
be exactly the same, only vertically shifted in the absolute value of
the emergent flux.  Indeed, the spectra are similar in the
long-wavelength (optical and IR) portion of the spectrum, but they
differ appreciably in the UV and EUV spectral region. In particular,
the Lyman edges of hydrogen and He~II change their appearance significantly.
Note that the non-LTE and LTE models for lower black hole masses in this
figure are very nearly the same.  This is expected, as the average density
$\rho\propto M/\dot M^2$ times some function of $r/r_g$ for radiation pressure
dominated annuli.  Hence for fixed
$\dot M/M^2$, $\rho$ scales as $M^{-3}$, implying that the lower black hole
mass models in figure \ref{FIGTOT4} have higher average densities and should
therefore be closer to LTE.

%
\vskip 2mm
\hbox{~}
\centerline{\psfig{file=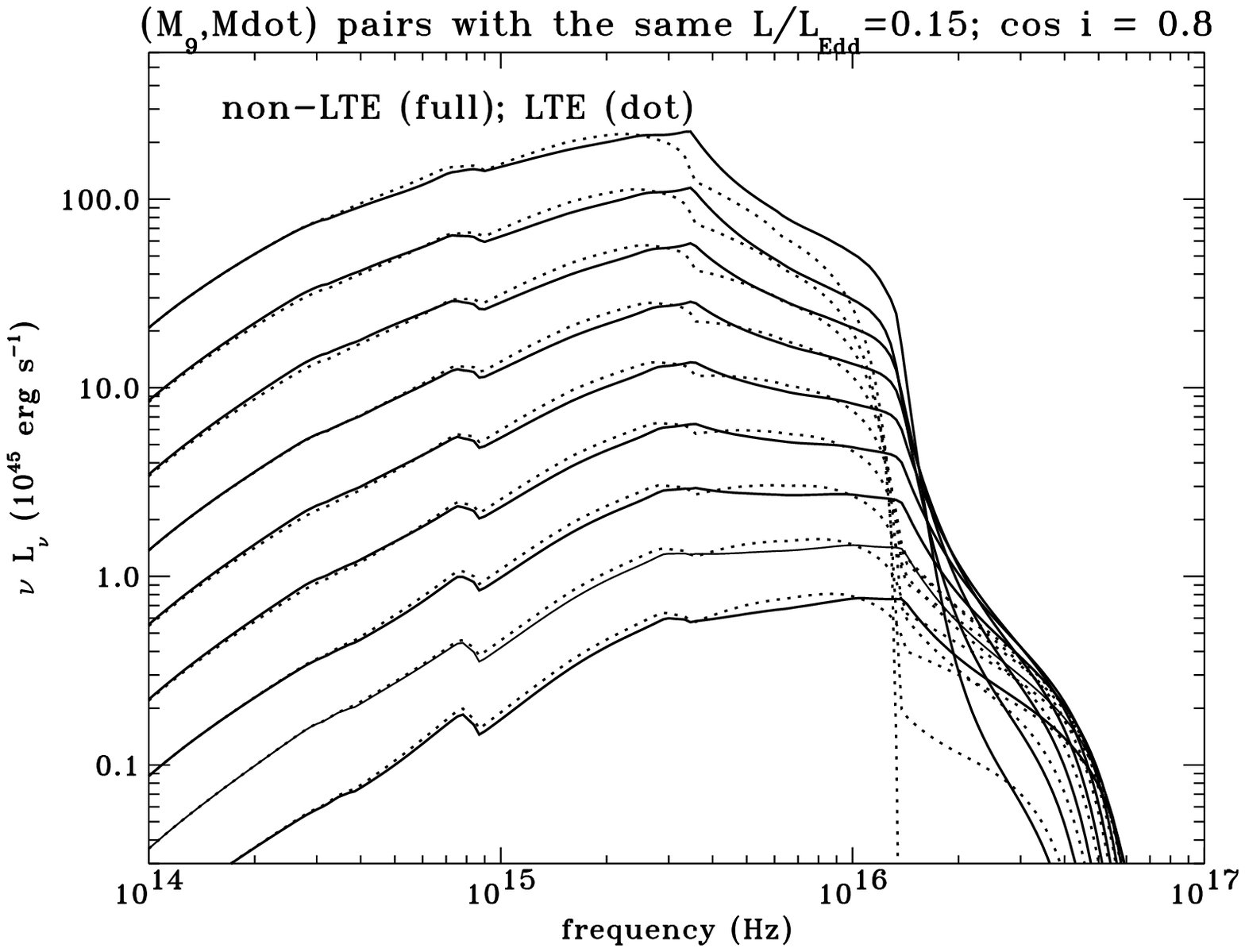,width=3.6in}} 
\noindent{
\scriptsize \addtolength{\baselineskip}{-3pt}
\figcaption{
Integrated spectral energy distribution $\nu L_\nu$  
(in erg s$^{-1}$) for the pairs of $(M_9,  \dot M)$ which have the
same $L/L_{\rm Edd}= 0.15$.
The numerical values of $M_9$ and $\dot M$ expressed in $M_\odot$ yr$^{-1}$,
are the same; the upper curve corresponds to the pair (32,32), and the lower
one to (1/8, 1/8).
Full lines: non-LTE models; dotted lines: LTE models.
The models with the highest $M_9$ (with the highest IR, optical, and UV
flux) have the largest He~II Lyman jump, and consequently the lowest flux
for $\nu > 2 \times 10^{16}$.
\label{FIGTOT5}
}
\vskip 3mm
\addtolength{\baselineskip}{3pt}
}
Finally, in figure \ref{FIGTOT5} we present a sequence of models with the
same $L/L_{\rm Edd}$. Since $L/L_{\rm Edd} \propto \dot M/M$, we
chose a sequence where $M_9$ and $\dot M$ (in $M_\odot$ yr$^{-1}$)
have the same value; the corresponding $L/L_{\rm Edd} = 0.15$.
Again, the shape of the spectrum is quite similar in the optical and IR regions,
while there is a progressively larger portion of EUV radiation for lower
black hole masses. The non-LTE effects are extreme for high-mass
holes in the He~II Lyman continuum, for which the LTE models predict
virtually zero flux.

\subsubsection{Effects of changing the viscosity parameter $\alpha$.}

In figure \ref{FIGAL01} we compare the predicted spectra for disk models
with $M_9$ =1 for two values of the viscosity parameter $\alpha$.
Although the spectra exhibit some differences, the effect of
different $\alpha$ is very small in the optical and UV region; the
only appreciable differences are found in the He II Lyman continuum.
This result is very encouraging because it shows that the effect of
the ad hoc viscosity parameter $\alpha$ is rather small, and therefore
the predicted spectra are not influenced significantly by this
inherent uncertainty.
Similar conclusions were reached in Paper II, albeit for a few
representative annuli.  
The sense of the net effect is that larger $\alpha$ leads to a lower density
and thus larger departures from LTE, which cause a somewhat higher flux in the 
He~II Lyman continuum.
%
\vskip 2mm
\hbox{~}
\centerline{\psfig{file=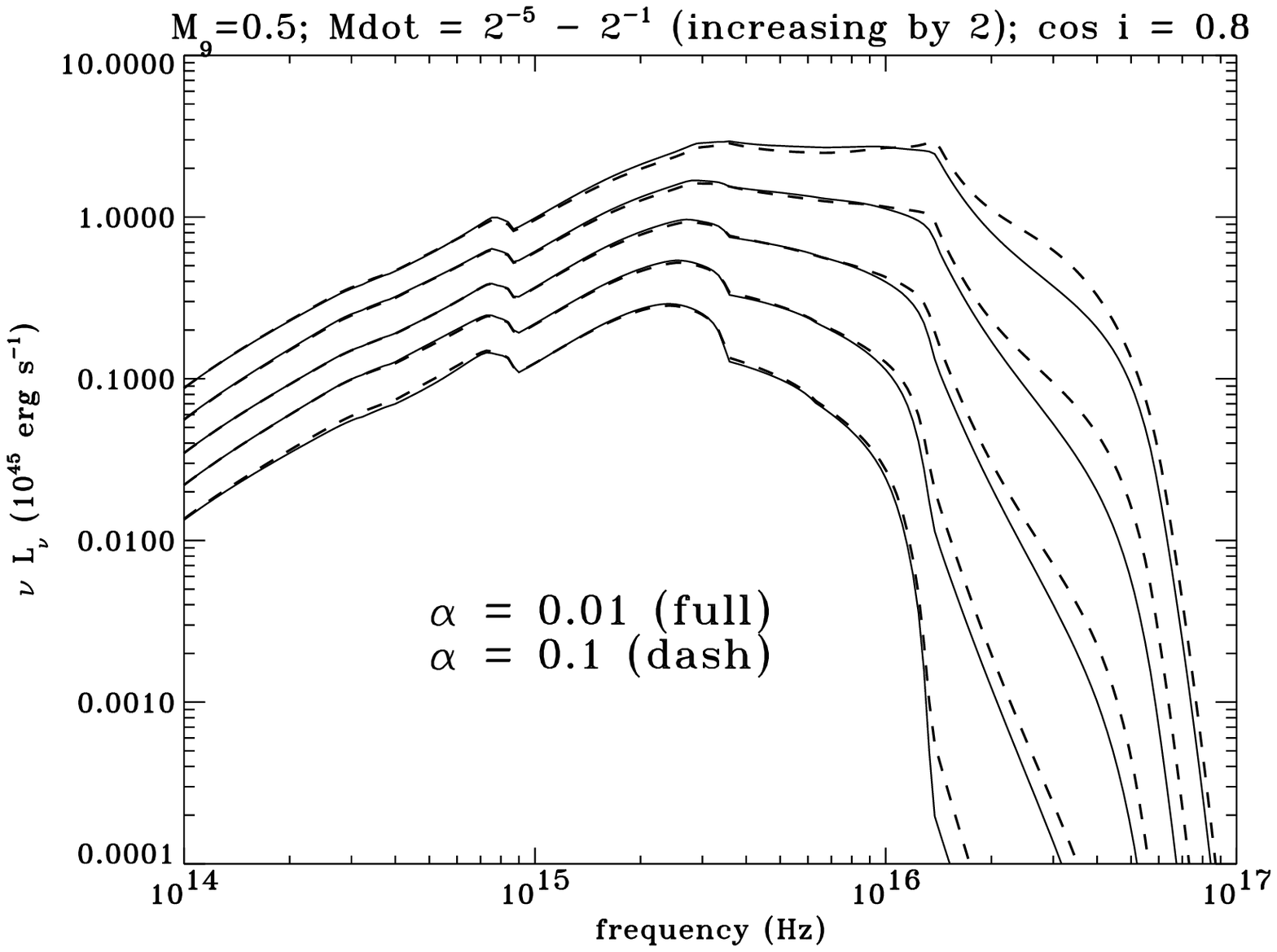,width=3.6in}} 
\noindent{
\scriptsize \addtolength{\baselineskip}{-3pt}
\vskip 1mm
\figcaption{
A comparison of predicted spectral energy distribution 
for models computed with $\alpha = 0.01$ (solid line), and with
$\alpha = 0.1$ (dashed lines), for models with $M_9 = 1/2$ and 
various values of $\dot M$.
\label{FIGAL01}
}
\vskip 3mm
\addtolength{\baselineskip}{3pt}
}

\subsubsection{Schwarzschild black holes}

%
\vskip 2mm
\hbox{~}
\centerline{\psfig{file=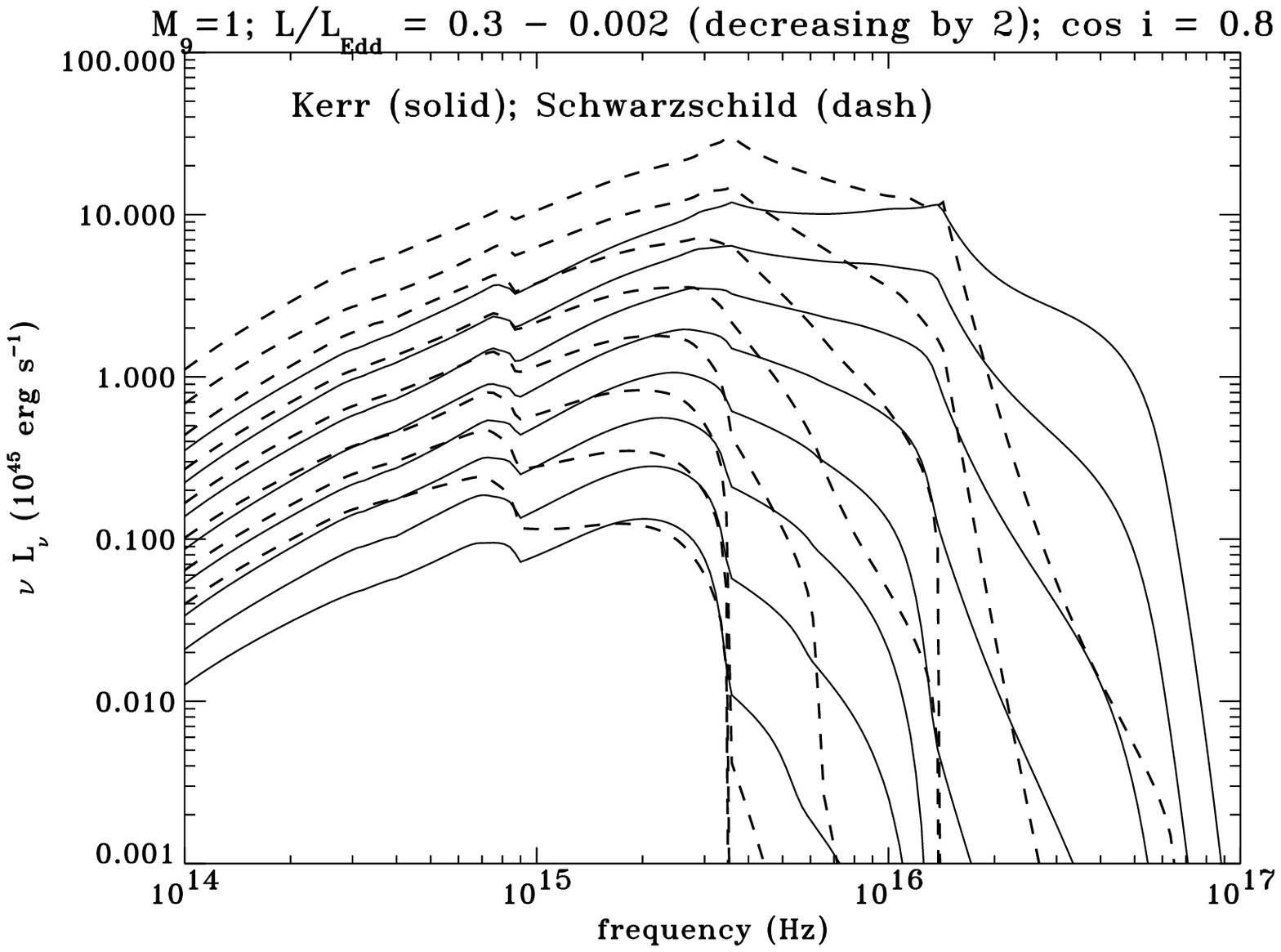,width=3.6in}} 
\noindent{
\scriptsize \addtolength{\baselineskip}{-3pt}
\vskip 1mm
\figcaption{
A comparison of predicted spectral energy distributions
for the sequence of models for a maximum rotating Kerr black hole (solid line), 
and for a Schwarzschild black hole (dashed lines), for models with $M_9 = 1$ 
and various values of $L/L_{\rm Edd}$.  The values of the mass accretion
rates are 2, 1, 1/2, 1/4, etc., for the Kerr hole, and the corresponding
values for the Schwarzschild hole are 5.613 times larger.
\label{FIGSCHW}
}
\vskip 3mm
\addtolength{\baselineskip}{3pt}
}
Finally, we present several representative spectra for the case of a
Schwarzschild black hole. Again, the full set of spectra is available
upon request.

In figure \ref{FIGSCHW} we compare the predicted spectra for disk models
with $M_9$ =1, for the maximum rotating Kerr and Schwarzschild 
black holes. The total luminosity of the corresponding pairs of models
is equal; the mass accretion rate is therefore a factor of 5.613 higher
for the Schwarzschild case to adjust for the different efficiencies.
The Kerr spectra tend to extend to higher frequencies for several
reasons all arising from the fact that their disks extend in to smaller
radii.  As a result, the maximum effective temperature found in
the disk is higher, and the relativistic effects strengthening the high
frequency spectrum away from the axis are also greater.

This completes our presentation of the overall integrated spectra of our
grid of models.  We now address some of their observational implications.

\subsection{Comparison with Other Models}

We first compare the spectrum of an LTE model with that computed by
Sincell \& Krolik (1998).  Figure \ref{FIGSK} shows a comparison between
one of their spectra and ours, computed for the same parameters.
The agreement is satisfactory, as the spectra differ by at most 20\% near
the peak.  The differences that do exist may be due to a number of
factors, ranging from technical numerical contrasts in the methods to
the detailed physical assumption made in both papers.

%
\vskip 2mm
\hbox{~}
\centerline{\psfig{file=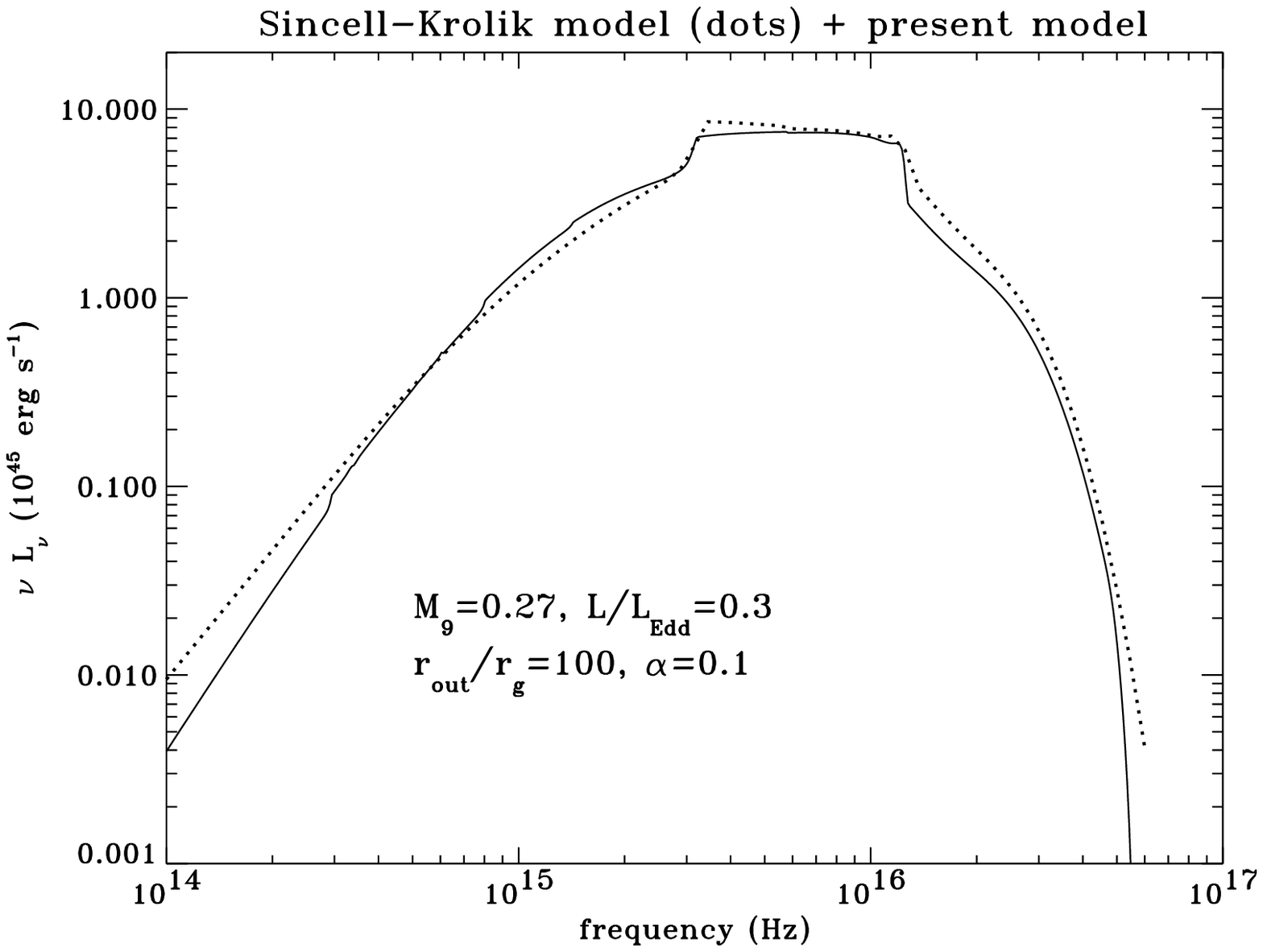,width=3.6in}} 
\noindent{
\scriptsize \addtolength{\baselineskip}{-3pt}
\vskip 1mm
\figcaption{
Comparison of an LTE disk model with
Sincell \& Krolik (1998).  Solid line is our model; dotted
line is theirs.  Model parameters are $M_9=0.27$, $a/M=0$,
$L/L_{\rm Edd}=0.3$,
$r_{out}=100r_g$, $\alpha=0.1$, and the disk is viewed face-on.
\label{FIGSK}
}
\vskip 3mm
\addtolength{\baselineskip}{3pt}
}

In figure \ref{FIGLAOR} we show a second comparison, this time to a Laor \&
Netzer (1989; Laor 1990) model, kindly supplied by A. Laor.
We choose a model with $M_9 = 1$,
$\dot M = 1$~M$_\odot$~yr$^{-1}$, (i.e., $L/L_{\rm Edd} = 0.15$), 
$\alpha = 0.01$, and
two values of $\cos i$: 0.8 and 0.2.  We have integrated the local
disk spectra out to a cutoff radius $r_{\rm out}/r_g = 217.8$
to agree with Laor's value.
The predicted spectra are generally similar, although there are
several interesting differencies.  Our models produce larger flux
in the immediate vicinity of the He II Lyman edge (due to non-LTE
effects leading to an emission edge for the hottest annuli), but a lower
flux for the highest frequencies (likely because Laor \& Netzer take
into account the effects of self-irradiation of the disk).
Our models produce lower flux in the UV and optical regions, which is a
consequence of the different vertical structure and of non-LTE effects.
Although the Laor \& Netzer models do not simply assume local blackbody
flux, figure \ref{FIGFLX2} is nevertheless quite indicative because it
shows that local blackbodies also produce a significantly larger
flux in the UV and optical region.
At IR wavelengths both models coincide because both use local
blackbody flux for the cool annuli.

%
\vskip 2mm
\hbox{~}
\centerline{\psfig{file=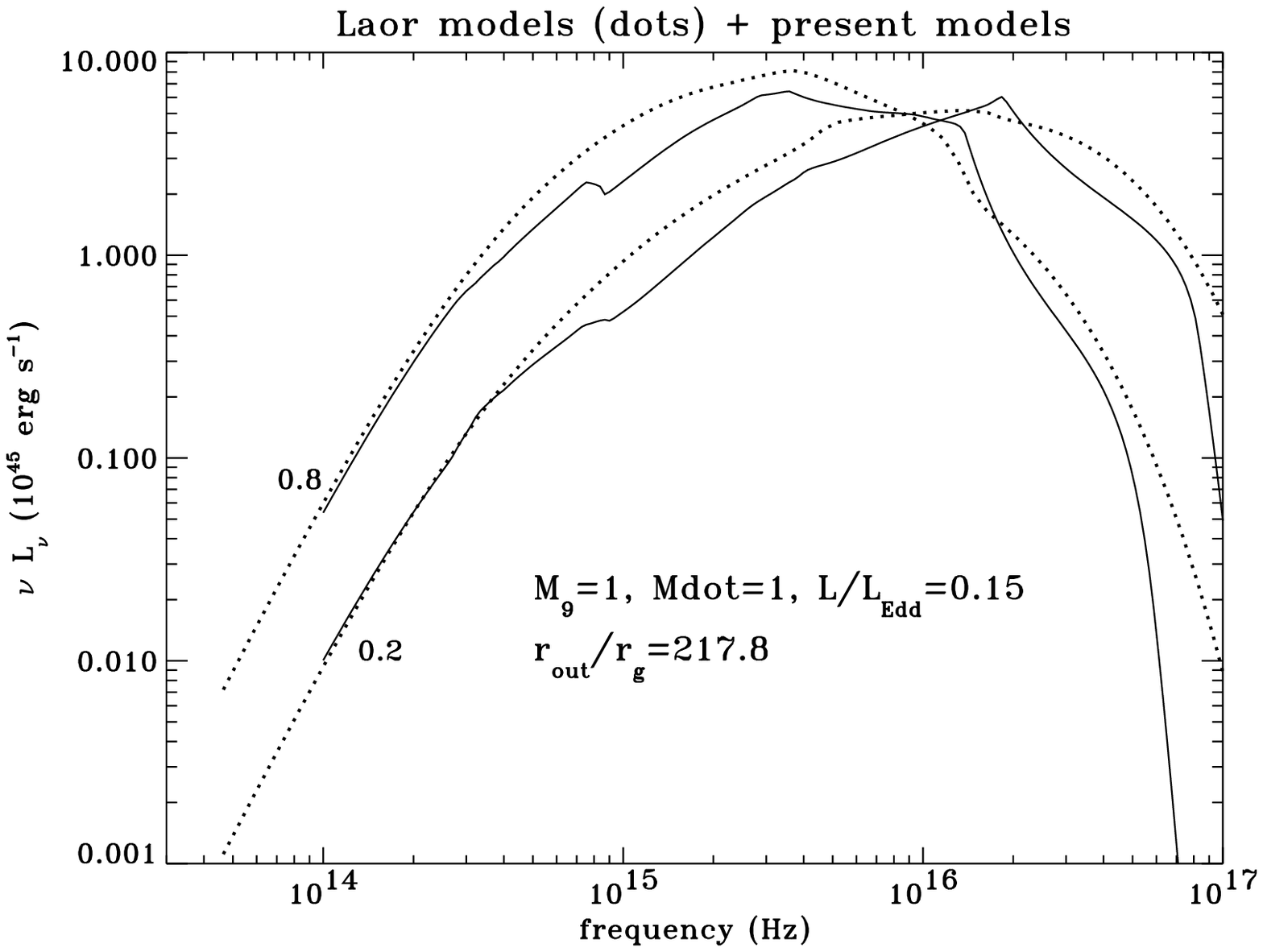,width=3.6in}} 
\noindent{
\scriptsize \addtolength{\baselineskip}{-3pt}
\vskip 1mm
\figcaption{
Comparison of a non-LTE disk model (solid curves) with
a Laor \& Netzer (1989; Laor 1990) model (dotted curves).
Model parameters are $M_9=1$, 
$L/L_{\rm Edd}=0.15$, $a/M=0.998$,
$r_{out}=217.8\, r_g$, $\alpha=0.01$, and $\cos i = 0.8$ and 0.2.
\label{FIGLAOR}
}
\vskip 3mm
\addtolength{\baselineskip}{3pt}
}

\subsection{Optical/Ultraviolet Colors}

A common criticism of accretion disk models is that if they are to produce
substantial ionizing photon flux, then they should have blue
optical/ultraviolet colors.  This is based on the long wavelength,
low frequency behavior of a blackbody disk, which has $F_\nu\propto\nu^\beta$,
with $\beta=1/3$.  We address the issue of ionizing photon flux in section
3.8 below, but here we wish to point out that our disks have quite red
optical/ultraviolet colors.  Figure \ref{FIGCOL} shows the logarithmic spectral
slope $\beta$ as measured between 1450${\rm\AA}$ and 5050${\rm\AA}$,
where $\beta$ is defined by the two corresponding frequencies by
\begin{equation}
{F_{\nu1}\over F_{\nu2}}=\left({\nu_1\over\nu_2}\right)^\beta.
\end{equation}
Our disk models have colors near the median value $\beta=-0.32$ for bright
quasars (Francis et al. 1991).  Indeed, even disks
with local blackbody emission have such red colors for these accretion rates
and masses (Koratkar \& Blaes 1999).  This is because the temperatures
are cool enough that the 1450--5050${\rm\AA}$ spectra are {\it not}
in fact in the long-wavelength limit.  Note that our model spectra
can be somewhat bluer than blackbody disks, but they are still
sufficiently red to explain the colors observed in bright quasars.

Figure \ref{FIGFRAN} shows some representative Kerr disk models viewed at
$i=37^\circ$ compared to the Francis et al. (1991) composite quasar spectrum.
The shorter wavelength composite spectrum of Zheng et al. (1997), scaled
to match the Francis et al. spectrum at 1285~${\rm\AA}$, is also shown.
This latter composite is thought to be more trustworthy than the Francis et
al. composite at wavelengths shorter than the Ly$\alpha$/NV line because of
corrections for intervening absorbers.
The models were chosen to have the right color based on figure \ref{FIGCOL},
and then a least squares fit was done to determine a single
multiplicative normalization factor.  The fit used supposedly line-free
continuum windows of the Francis et al. (1991) composite spectrum, as defined
by their figure 7: 1283-1289${\rm\AA}$, 1321-1329${\rm\AA}$,
1455-1475${\rm\AA}$, 2196-2208${\rm\AA}$, 2236-2246${\rm\AA}$,
3024-3036${\rm\AA}$, 3928-3936${\rm\AA}$, 4035-4045${\rm\AA}$,
4150-4220${\rm\AA}$, and 5464-5476${\rm\AA}$.  Note that the models
shown have $\dot M/M^2$ ranging from 1/16 to 1/2, with the lower mass models
having the smaller values of this quantity.  This is consistent with the
behavior shown in figure \ref{FIGTOT4}: models with fixed $\dot M/M^2$ have
slightly bluer optical/UV colors for decreasing black hole mass.
%
\vskip 2mm
\hbox{~}
\centerline{\psfig{file=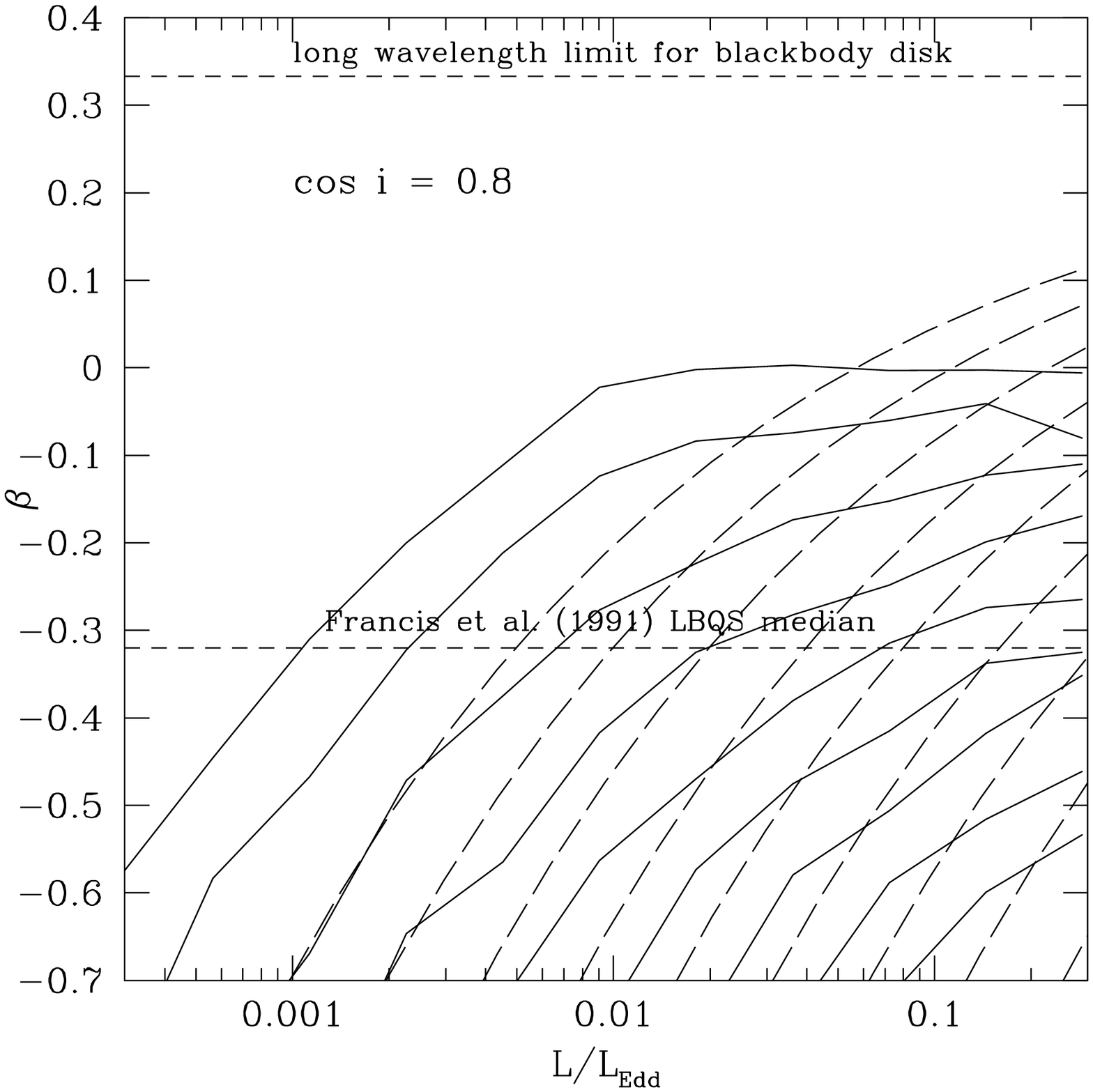,width=3.6in}} 
\noindent{
\scriptsize \addtolength{\baselineskip}{-3pt}
\vskip 1mm
\figcaption{
Optical/ultraviolet spectral slope between 1450${\rm\AA}$
and 5050${\rm\AA}$ for Kerr disk models with 37$^\circ$ viewing angle.
Solid curves show our models with viscosity parameter $\alpha=0.01$, while
the long dashed curves show local blackbody models.  From top to bottom, the
curves are for the nine increasing black hole masses of our grid:
$M_9 = 1/8, 1/4, 1/2, 1, 2, 4, 8, 16, 32$.
\label{FIGCOL}
}
\vskip 3mm
\addtolength{\baselineskip}{3pt}
}

These fits demonstrate
that, while it is easy to recover the overall red color of the Francis et al.
(1991) composite spectrum, explaining the shorter wavelength far ultraviolet
emission seen in the Zheng et al. (1997) composite is more problematic.  Two
of the models shown in figure \ref{FIGFRAN} do in fact bracket the Zheng et
al. composite, but they turn over at the shortest wavelengths shown in
the figure.  This might be a problem in view of the fact that an extrapolation
of the Zheng et al. composite joins up with the {\it ROSAT} soft X-ray
composite of Laor et al. (1997), implying that there is no cutoff.
However, it is also conceivable that some other choice of model parameters
(disk inclination, black hole spin) might work
better.  Notice also that Zheng et al. suggested that in order to explain the
Lyman continuum flux one has to invoke the presence of a Comptonizing
corona with temperature about $4 \times 10^8$ K with optical depth
of the order of unity.  In any case, composite spectra made from many 
sources may of course be unphysical,
but these fits probably give some indication of how our models will fare in
explaining data from individual sources.

It is also noteworthy that the low luminosity models shown in figure
\ref{FIGFRAN} have rather strong spectral features in the region of the
hydrogen Lyman limit, which are generally
not observed in quasars.  However, the models with the lowest luminosities
are probably not representative of the quasars that make up the composite
spectra.
We now address this important issue of Lyman edges in AGN.

%
\vskip 2mm
\hbox{~}
\centerline{\psfig{file=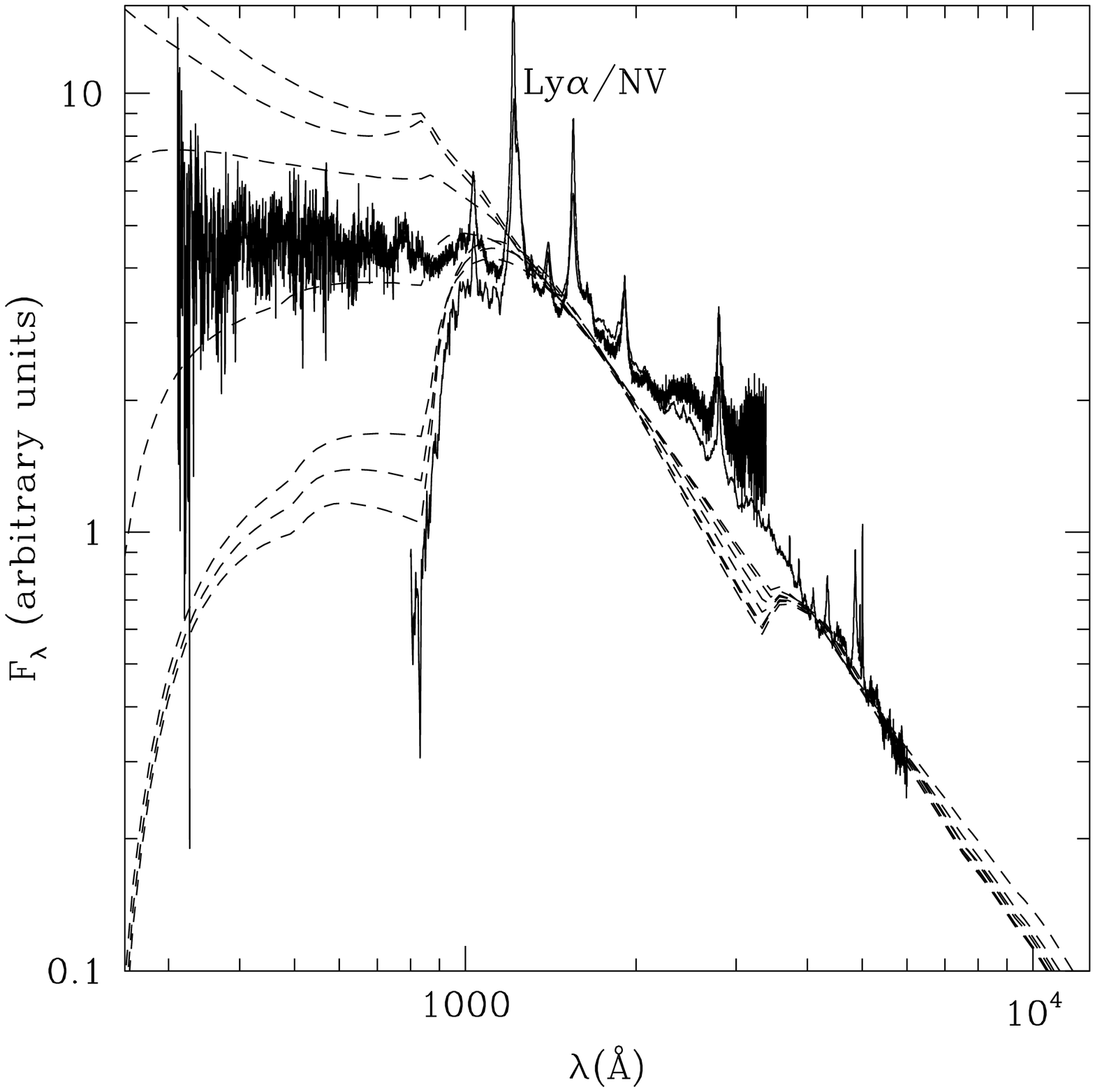,width=3.6in}} 
\noindent{
\scriptsize \addtolength{\baselineskip}{-3pt}
\vskip 1mm
\figcaption{
Representative fits of some Kerr disk models with viscosity
parameter $\alpha=0.01$ and $\cos i=0.8$ to the Francis et al. (1991)
composite quasar spectrum (solid line).  The shorter wavelength composite
spectrum of Zheng et al. (1997) is also shown (solid line with more noise).
From top to bottom, the dashed
lines correspond to models with $(M_9,\dot M)=$ (8,16), (4,8), (2,1), (1,1/8),
(0.5,1/64), (1/4,1/256), and (1/8,1/1024), respectively.
\label{FIGFRAN}
}
\vskip 3mm
\addtolength{\baselineskip}{3pt}
}

\subsection{The Lyman Edge Region}

Quasars and active galactic nuclei are observed to have almost no intrinsic
spectral features near the hydrogen Lyman limit (e.g. Antonucci, Kinney, \&
Ford 1989; Koratkar, Kinney, \& Bohlin 1992), and this has been a
longstanding problem with accretion disk models (e.g. Krolik 1999a, Koratkar
\& Blaes 1999).
To quantify the strength of Lyman edge features in our model disk spectra,
we calculate the relative change in flux at $\pm50{\rm \AA}$ across the edge
according to
\begin{equation}
{\Delta F_\lambda\over F_\lambda}\equiv{F_{962{\rm \AA}}-F_{862{\rm \AA}}\over
{\rm min}[F_{962{\rm \AA}},\, F_{862{\rm \AA}}]}.
\end{equation}
Hence $\Delta F_\lambda/F_\lambda$ is positive for an absorption edge and
negative for an emission edge.  Figure \ref{LYMAN1} shows the variation of
this quantity with accretion luminosity and disk inclination angle for all
our models around a $M_9=1$ Kerr hole.  Figure \ref{LYMAN2} shows the
same thing but for a fixed inclination angle of 37$^\circ$ and varying black
hole masses.  As was already apparent from the overall spectra shown in
previous sections, substantial Lyman absorption edges are present in all
our low luminosity disk models at modest, near face-on viewing angles.
However, high luminosity models have greatly reduced edges, particularly
for the lower mass black holes.  Because of their lower effective temperatures,
higher mass black holes require higher Eddington ratios before the
absorption edges are removed.
%
\vskip 1mm
\hbox{~}
\centerline{\psfig{file=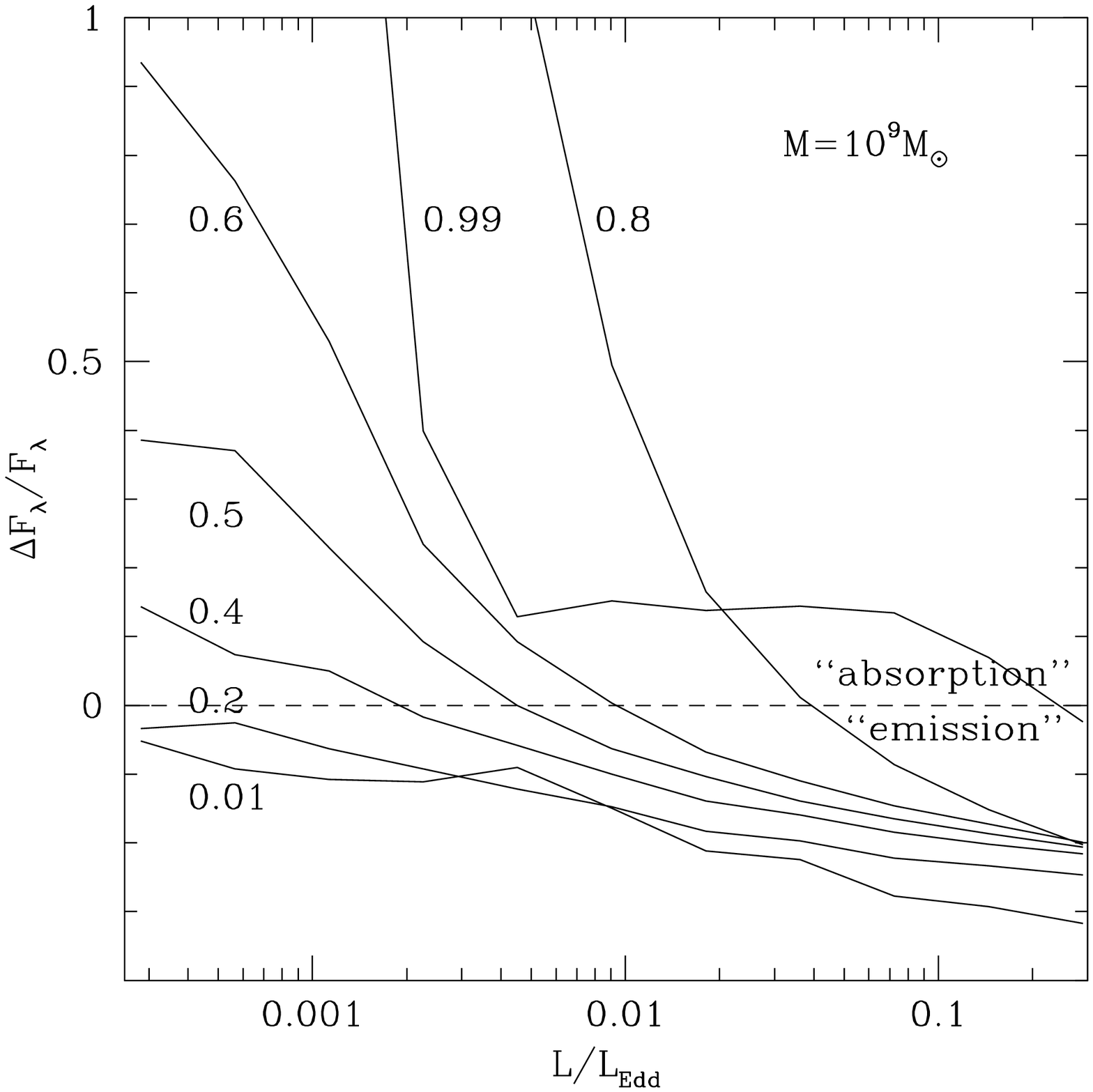,width=3.6in}} 
\noindent{
\scriptsize \addtolength{\baselineskip}{-3pt}
\figcaption{
Measure of the hydrogen Lyman edge strength as a function of
accretion luminosity, in units of the Eddington luminosity, for $\alpha=0.01$
accretion disks around Kerr holes with mass $M_9=1$.  The curves are labeled
with the value of $\cos i$.
\label{LYMAN1}
}
\vskip 1mm
\addtolength{\baselineskip}{3pt}
}
It is important to emphasize that the absorption and emission edges can be
smeared out by the varying Doppler shifts and gravitational redshifts
in the accretion flow around the black hole when the viewing direction is
at least somewhat off-axis.  This is illustrated in figures
\ref{LYMAN3} and \ref{LYMAN4}, 
which show the actual spectral energy distributions of
some of our models in the Lyman limit region.  The high luminosity models,
which are probably most
relevant for observed quasars, show very little in terms of sharp changes
in flux, and our edge strength parameter shown in figures 
\ref{LYMAN1} and \ref{LYMAN2} really
reflects an overall spectral slope, {\it not} an emission edge, in this
wavelength region.  These high luminosity models are still capable of
explaining the observed red colors of quasars, as shown in figures
\ref{FIGCOL} and \ref{FIGFRAN}.

%
\vskip 1mm
\hbox{~}
\centerline{\psfig{file=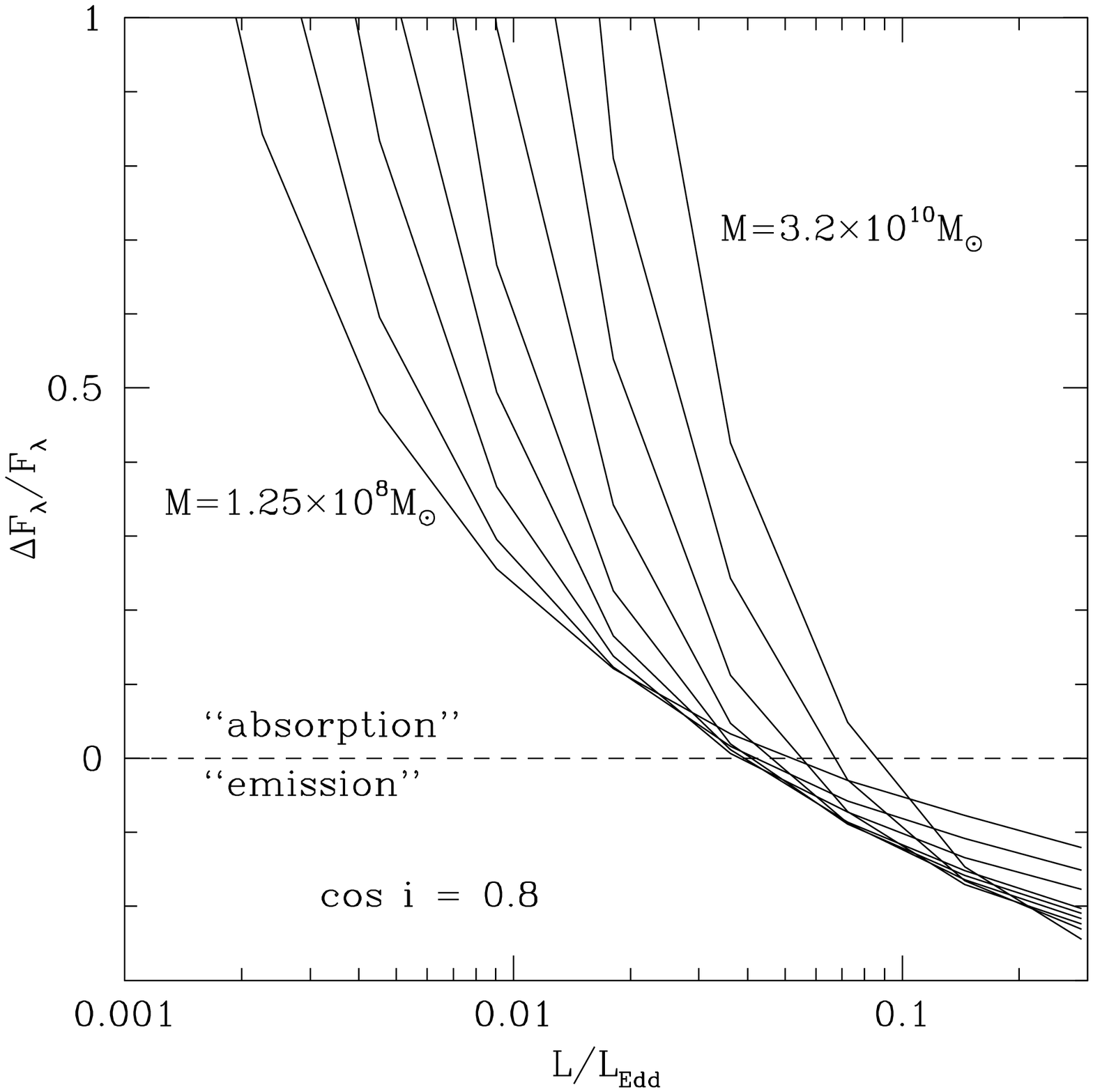,width=3.6in}} 
\noindent{
\scriptsize \addtolength{\baselineskip}{-3pt}
\figcaption{
Same as figure \ref{LYMAN1} but for fixed inclination angle of
$37^\circ$ and varying Kerr black hole mass.
\label{LYMAN2}
}
\addtolength{\baselineskip}{3pt}
}
%

%
\vskip 2mm
\hbox{~}
\centerline{\psfig{file=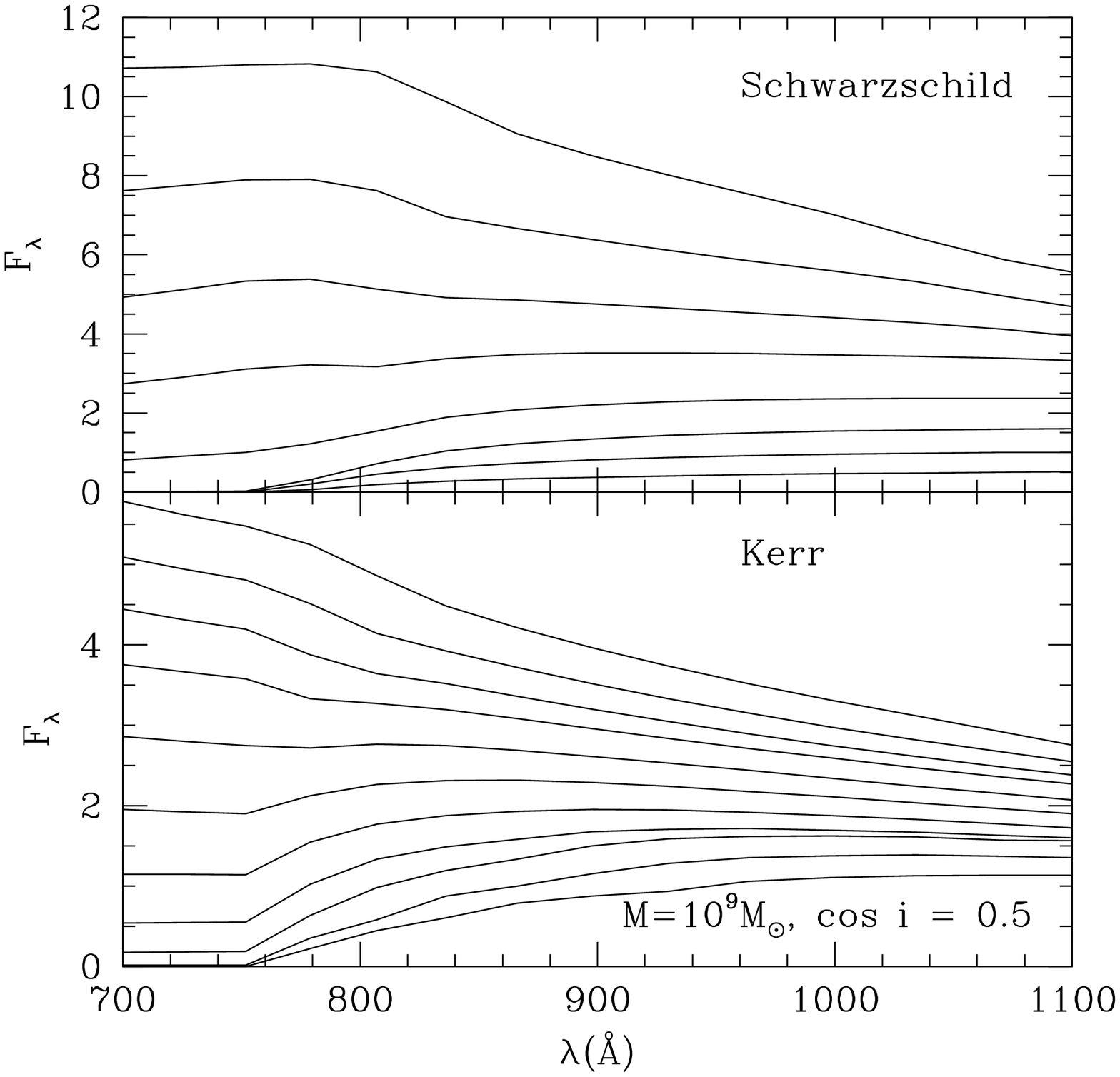,width=3.6in}} 
\noindent{
\scriptsize \addtolength{\baselineskip}{-3pt}
\vskip 1mm
\figcaption{
Spectral energy distributions in wavelength near the hydrogen
Lyman limit for fixed Schwarzschild and Kerr black hole mass of 
$M_9=1$ and viewing
angle of 60$^\circ$.  From bottom to top, the curves represent accretion
rates of
$2^{-9}, 2^{-8},\ldots, 2^0, 2^1$~M$_\odot$~yr$^{-1}$, 
respectively, except in
the Schwarzschild case where we have dropped the three lowest accretion rates.
(The $2^{-6}$~M$_\odot$~yr$^{-1}$ Kerr curve is the same as the curve with the
strongest Lyman absorption edge shown in figure \ref{FIGFRAN}.)
Each curve has had its flux multiplied
by a different constant factor in order to maximize clarity with all the
curves on the same plot.  All models shown here have a viscosity parameter
$\alpha=0.01$.
\label{LYMAN3}
}
\vskip 3mm
\addtolength{\baselineskip}{3pt}
}

Reduction of the Lyman edge feature is caused both by relativity and
by summing over emission and absorption edges from the individual annuli at
different radii.  Only relativity (i.e. Doppler shifts) can smear out a flux
discontinuity, however.  We have tried integrating
spectra {\it without} the relativistic transfer function for the
$M=10^9$~M$_\odot$, $\dot M=1$~M$_\odot$~yr$^{-1}$, $\cos i=0.8$ case and
found that without relativity, a substantial emission
edge discontinuity exists in the integrated spectrum at 912~${\rm\AA}$.
Relativity is not sufficient to smear out absorption edges in the low
luminosity models, because the absorption edges are very strong in the
spectra of all annuli in such models.  Instead, the edges are simply blue
shifted or red shifted away from 912~${\rm\AA}$.

%
\vskip 2mm
\hbox{~}
\centerline{\psfig{file=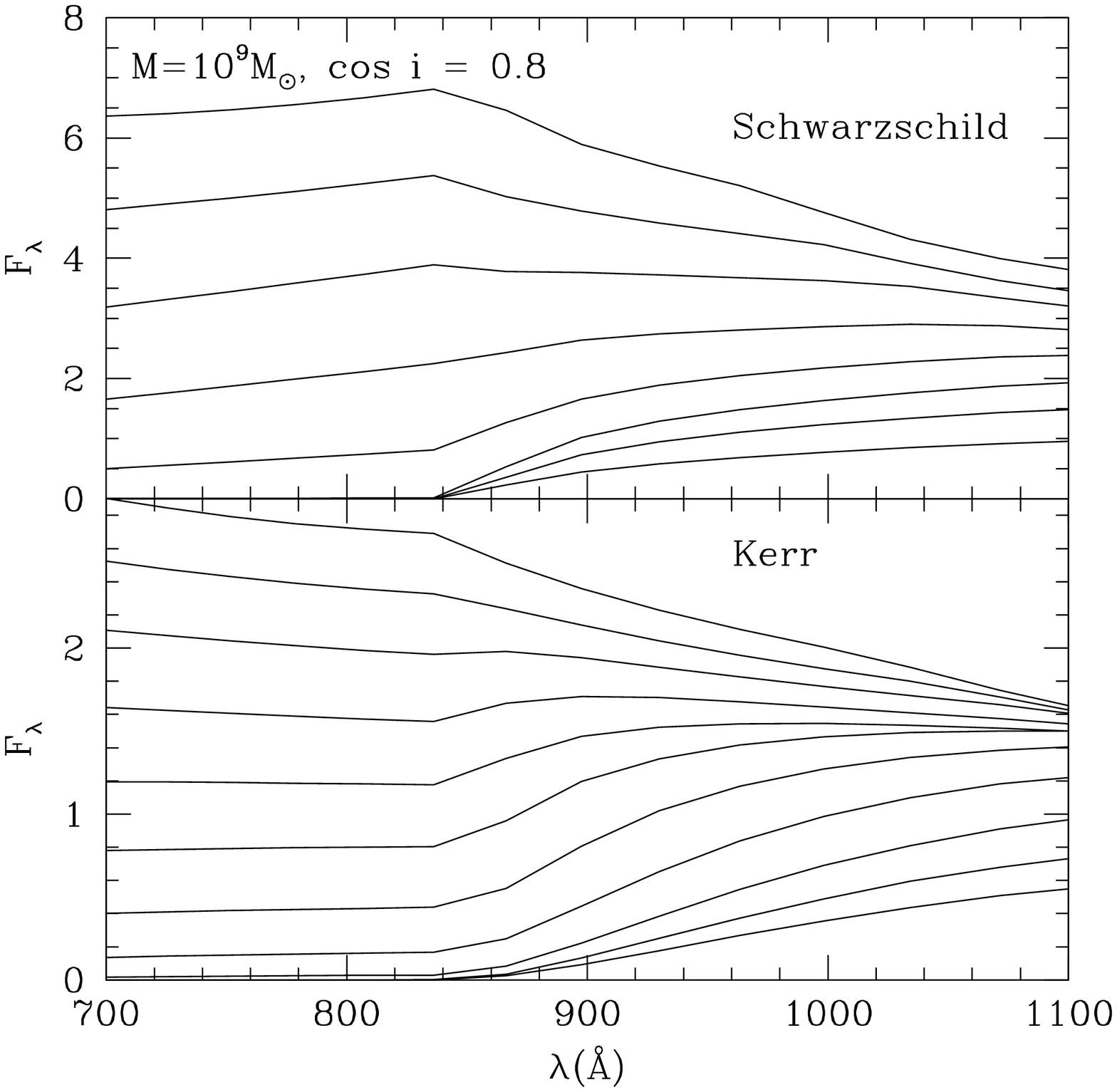,width=3.6in}} 
\noindent{
\scriptsize \addtolength{\baselineskip}{-3pt}
\vskip 1mm
\figcaption{
Same as figure \ref{LYMAN3} but for viewing angle of $37^\circ$.
\label{LYMAN4}
}
\vskip 3mm
\addtolength{\baselineskip}{3pt}
}
Sharp changes in slope are
present in the low luminosity Kerr spectra near $\simeq750{\rm \AA}$ for
$\cos i=0.5$, and in most of the spectra near $\simeq840{\rm \AA}$ for
$\cos i=0.8$.  In both cases these are at substantially shorter wavelengths
than the Lyman limit because of Doppler blue shifts, but they
might nevertheless be observable in quasar spectra.  From our models, we have
calculated the maximum local change in logarithmic slope in the region
812-1012~\AA~, and the results are illustrated in figure \ref{LYMSLOPE} for a
viewing angle of 37$^\circ$.  We choose a sign convention such that
a positive value of the slope change indicates a spectrum that becomes
steeper as the frequency increases, i.e., positive slope change is associated
with absorption at the edge.  All models display a similar behavior at this
viewing angle, so we only discuss the $M=10^9$~M$_\odot$ case in detail
(cf. figure \ref{LYMAN4} and the bold curve in figure \ref{LYMSLOPE}).  
At both low and high accretion rates, the maximum slope change occurs at
$\simeq840{\rm \AA}$ for this black hole mass.   As the accretion rate
diminishes, $\Delta(d\ln F_\lambda/d\ln\lambda)$ grows, reflecting an
increasingly stronger smeared absorption edge.  The reverse
is true at high accretion rates.  At intermediate accretion rates, the
maximum slope discontinuity shifts to $\simeq900{\rm \AA}$, where the
spectra change from positive to negative slopes, reflecting a slight maximum
in the flux around this wavelength for these models.

In conclusion, the disk-integrated theoretical spectra for high 
Eddington ratio ($L/L_{\rm Edd}$) disks
do not show significant features at the Lyman
edge at 912 \AA, for both Kerr and Schwarzschild black holes.  The only
associated feature is a change of the slope of the Lyman continuum, blue
shifted by $\sim100 - 200$~\AA,
depending on the inclination, mass, and to some extent on the black hole spin.
It is likely that even this feature will be affected by additional physics,
particularly metal line blanketing, which we will address in a future paper.
%
\vskip 2mm
\hbox{~}
\centerline{\psfig{file=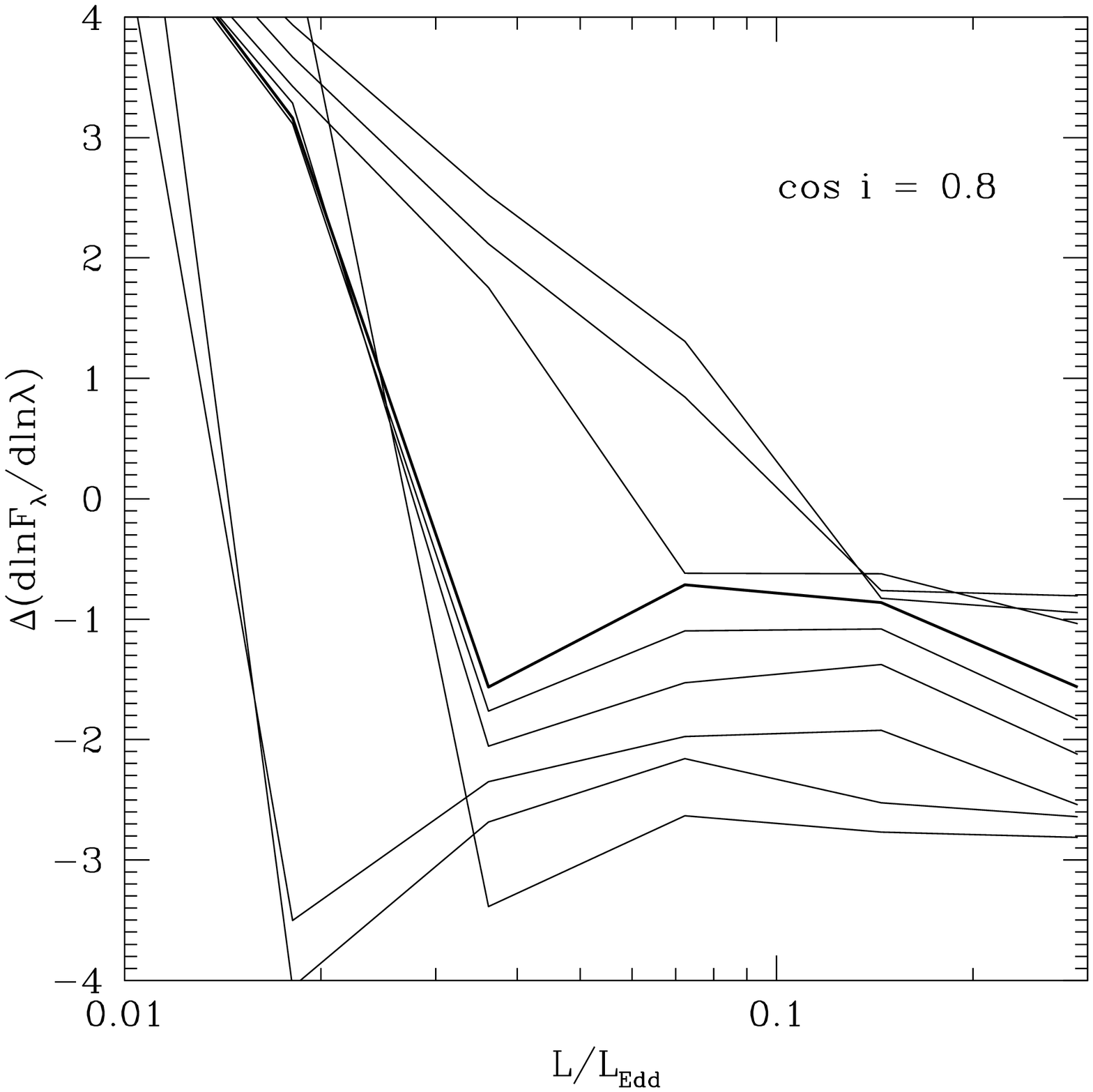,width=3.6in}} 
\noindent{
\scriptsize \addtolength{\baselineskip}{-3pt}
\vskip 1mm
\figcaption{
Maximum local change in logarithmic slope in 812-1012~\AA~
region, as a function of accretion rate,
for $\alpha=0.01$ accretion disks viewed at $37^\circ$ around Kerr holes of
masses $(2^{-3}, 2^{-2}, \ldots, 2^4, 2^5)\times10^9$~M$_\odot$.  The bold
curve is the $M=10^9$~M$_\odot$ case illustrated in the bottom half of
figure \ref{LYMAN4}.
\label{LYMSLOPE}
}
\vskip 3mm
\addtolength{\baselineskip}{3pt}
}

\subsection{Polarization}

In addition to computing spectra, we have also calculated the polarization
in our complete grid of models, and this information is also available on
request.  Once again, no ad hoc assumptions are made here: the 
polarization is computed exactly in the radiative transfer calculation.
In order to keep the parameter space as
simple as possible, we do not include the effects of Faraday rotation
by magnetic fields in the photosphere of the accretion disk, which can be
important in modifying the polarization signature (e.g. Agol, Blaes, \&
Ionescu-Zanetti 1998).

Figure \ref{POLGEN} shows the degree and position angle of the polarization
for various inclination angles for our $\alpha=0.01$,
$\dot M=1$~M$_\odot$~yr$^{-1}$ disk models around a $M=10^9$~M$_\odot$ Kerr
hole.  These results are quite similar to those of Laor, Netzer, \&
Piran (1990).  In particular, the plane of polarization is parallel to
the disk plane at optical/UV frequencies, but rotates at higher frequencies
due to general relativistic effects.  Our predicted polarizations are
higher than those of Laor et al. (1990), and our polarization generally
dips redward and rises blueward of each continuum edge (cf. especially
the hydrogen Balmer edge and He~II Lyman edge).  These differences are
due largely to our more careful treatment of the vertical structure,
the radiation field anisotropy, and the overall effects of absorption
opacity.  

Polarization near the Lyman limit of hydrogen has
produced considerable recent interest due to the observation of steep rises
in some quasars (Impey et al. 1995, Koratkar et al. 1995, Koratkar et al.
1998).  For illustration purposes, we show the degree of polarization
at a viewing angle of 60$^\circ$ predicted by our models for 
$M_9=1$
black holes in figure~\ref{POL}.  
This figure should be compared to figure \ref{LYMAN3},
which shows the corresponding spectra.  Cooler disk models generally produce
large polarizations, even larger than that for a pure electron scattering
atmosphere (2.25\% at this viewing angle, Chandrasekhar 1960).  The reason
for this is the enhanced anisotropy (limb darkening) of the radiation
field due to vertical thermal source function gradients (Blaes \& Agol 1996).
However, steep rises in polarization, and a steep rise in polarized flux as
observed in PG 1630+377 (Koratkar et al. 1995), are not produced by our
models.  This is partly due to the smearing and rotation of the plane of
polarization by the relativistic transfer function (cf.  Shields, Wobus,
\& Husfeld 1998).

%
\vskip 2mm
\hbox{~}
\centerline{\psfig{file=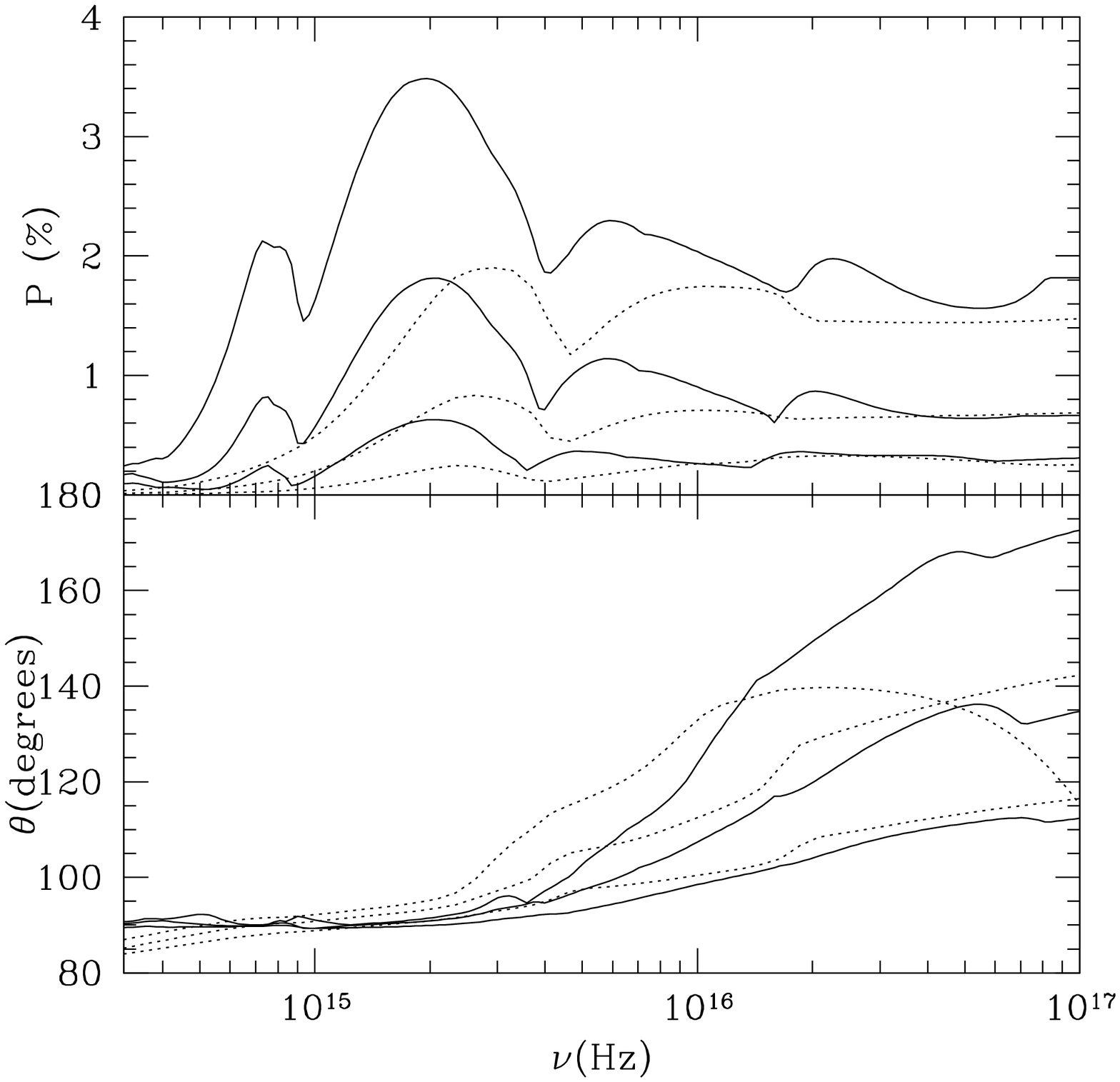,width=3.6in}} 
\noindent{
\scriptsize \addtolength{\baselineskip}{-3pt}
\vskip 1mm
\figcaption{
Polarization degree and position angle ($\theta$) for
$\alpha=0.01$,
$\dot M=1$~M$_\odot$~yr$^{-1}$ disks around a $M=10^9$~M$_\odot$ Kerr
hole, for different viewing angles.  Solid curves correspond to our models,
while dotted curves are those of Laor, Netzer, \& Piran (1990).  From top
to bottom at high frequencies in the degree of polarization figure,
the different curves correspond to $\cos i=0.2$, 0.5, and 0.8,
respectively.  This order is reversed (i.e. bottom to top) in the
position angle figure.  A position
angle of $\theta=90^\circ$ corresponds to the plane of polarization being
parallel to the disk plane.
\label{POLGEN}
}
\vskip 3mm
\addtolength{\baselineskip}{3pt}
}
The optical/UV polarization shown in figure \ref{POLGEN} (in degree,
position angle, and wavelength dependence) is generally not observed in 
AGN and quasars (Berriman et al. 1990, Antonucci et al. 1996).
Once again, it is likely that the polarization of the radiation field will be
affected by additional physics.  In addition to Faraday rotation (which
usually suppresses polarization), the
additional absorption opacity from metal line blanketing in this region of
the spectrum will probably reduce the polarization.  Dust and electron
scattering at larger distances from the continuum source can also modify
the polarization signature (Kartje 1995).

%
\vskip 2mm
\hbox{~}
\centerline{\psfig{file=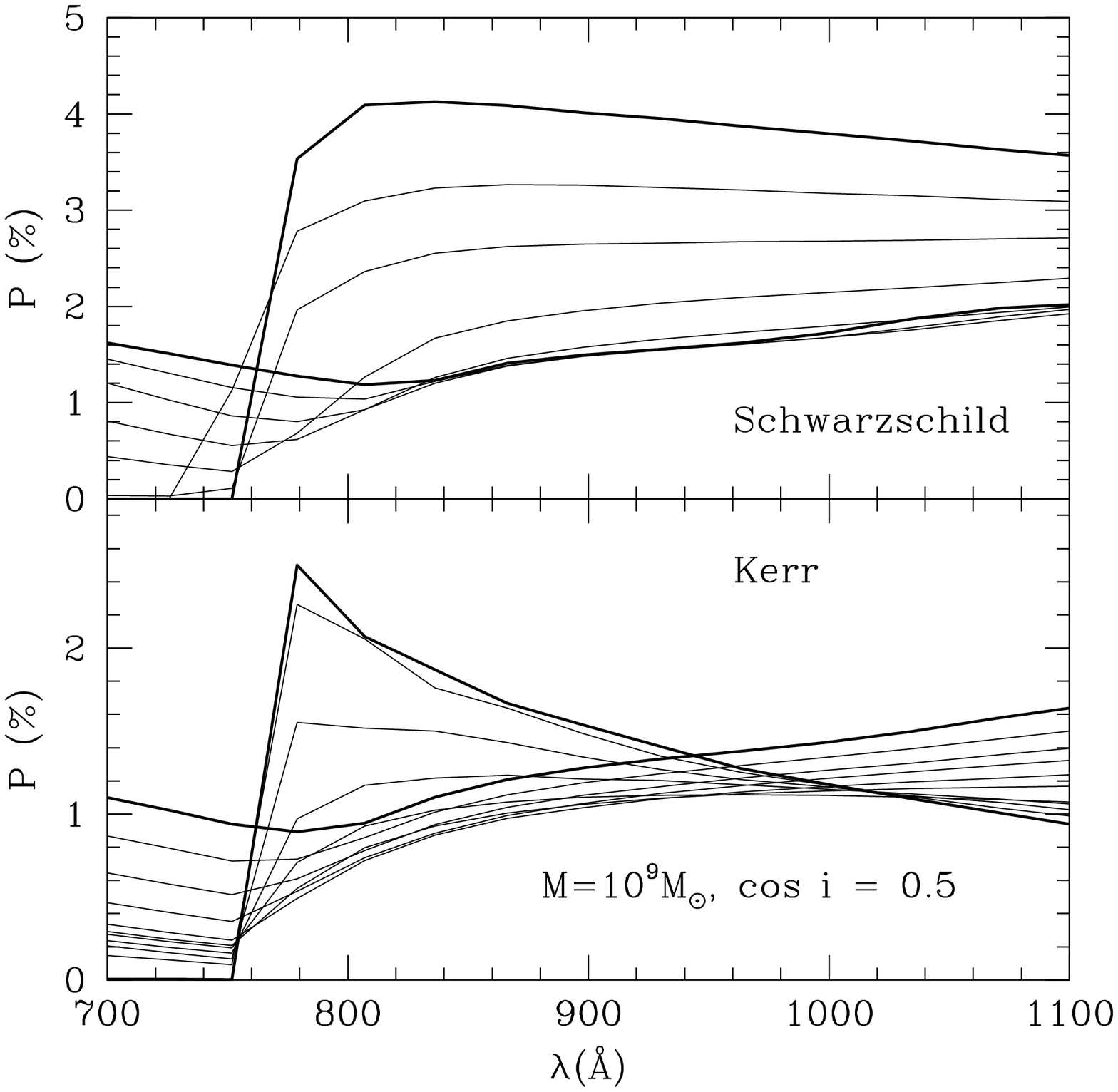,width=3.6in}} 
\noindent{
\scriptsize \addtolength{\baselineskip}{-3pt}
\vskip 1mm
\figcaption{
Degree of polarization in the hydrogen Lyman limit region for
disks around $10^9$~M$_\odot$ Schwarzschild and Kerr black holes and a
viewing angle of 60$^\circ$.  The models shown here are the same as those
plotted in figure \ref{LYMAN3}, with the bold curves in each plot representing
the highest and lowest accretion rates.  (The lowest accretion rate gives
the highest polarization around 800${\rm \AA}$.)
\label{POL}
}

\vskip 3mm
\addtolength{\baselineskip}{3pt}
}

\subsection{Ionizing Continua}

The radiation from accretion disks is often thought to supply most of the
photoionizing
continuum for the broad and narrow line regions of AGN.  In view of this
important application, we present the number of photons in the
HI and HeII ionizing continua for disks with a range of accretion rates and
inclination angles around $10^9$~M$_\odot$ Kerr holes in figures \ref{ION} and
\ref{IONHE}.  For comparison, we also show the ionizing continua for the 
corresponding
models with local blackbody emission.  For this particular black hole mass,
our models generally predict somewhat fewer ionizing photons in the hydrogen
Lyman continuum than the corresponding blackbody disks.  The exception is
for near face-on disks at high luminosities.  The reason is that at each
annulus in the disk, our models generally have fewer low energy photons and
more high energy photons compared to a blackbody at the same effective
temperature (cf. figure \ref{FIGFLX2} and section 3.1 above).  
This can reduce the hydrogen Lyman
continuum, while at the same time increasing the HeII Lyman continuum.
Indeed, figure \ref{IONHE} shows that, except at low luminosities, 
our disk models
generally produce more HeII Lyman continuum photons than local blackbody
disks.  The reason for the dearth of photons at low luminosities compared
to blackbody models is the strong absorption edges present in
these models.

%
\hbox{~}
\centerline{\psfig{file=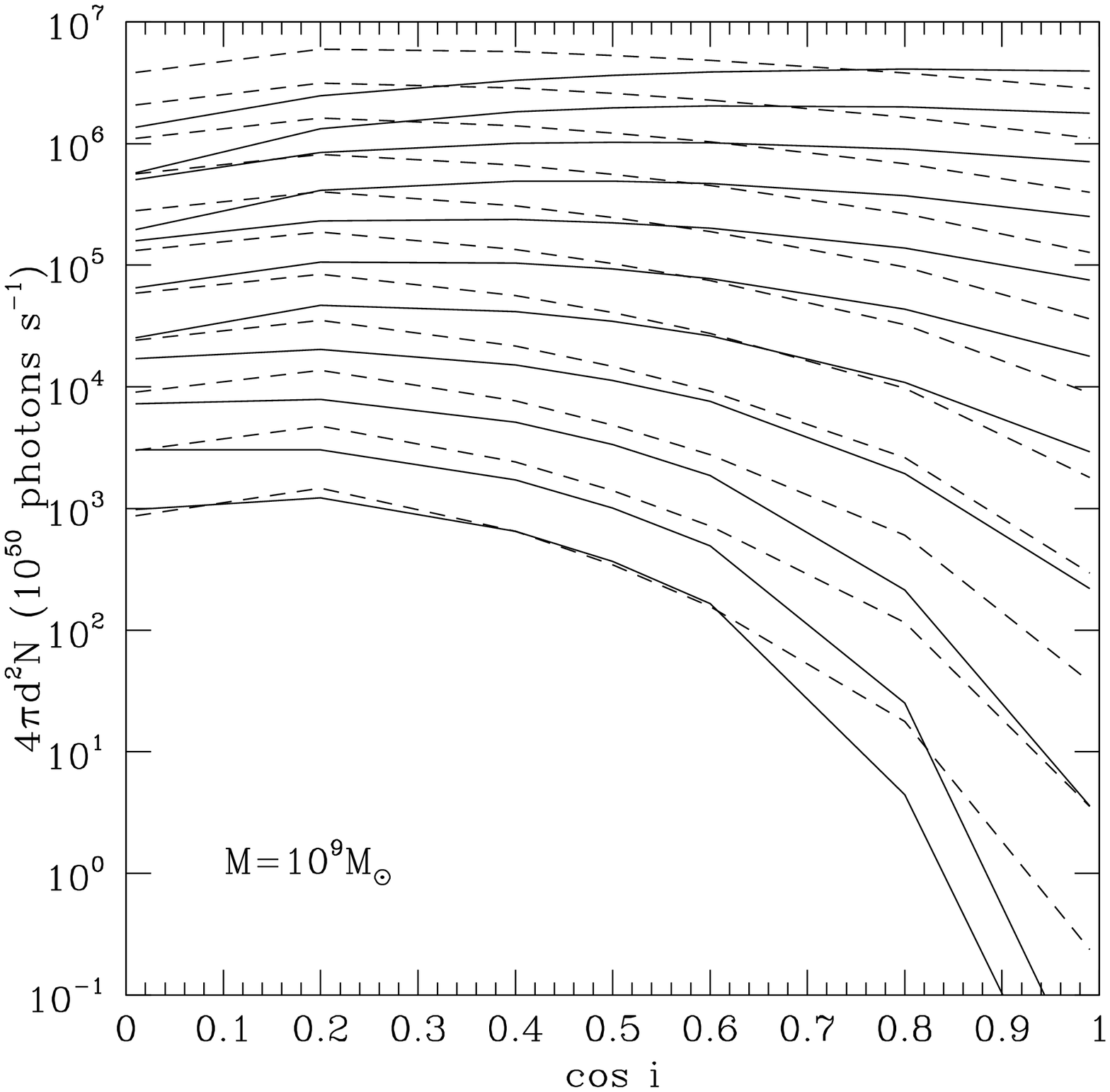,width=3.5in}} 
\noindent{
\scriptsize \addtolength{\baselineskip}{-3pt}
\figcaption{
Hydrogen Lyman continuum photon flux (times $4\pi d^2$, where $d$
is the distance
to the quasar) for $\alpha=0.01$ disks around a 
$M_9=1$
Kerr black hole, as
a function of viewing angle $i$.  Solid and dashed curves correspond
to our non-LTE models and local blackbody disks, respectively.  From top to
bottom, the curves correspond to accretion rates of
$2^1, 2^0, \ldots, 2^{-8}, 2^{-9}$~M$_\odot$~yr$^{-1}$,
respectively.
\label{ION}
}
\addtolength{\baselineskip}{3pt}
}
%
%
\hbox{~}
\centerline{\psfig{file=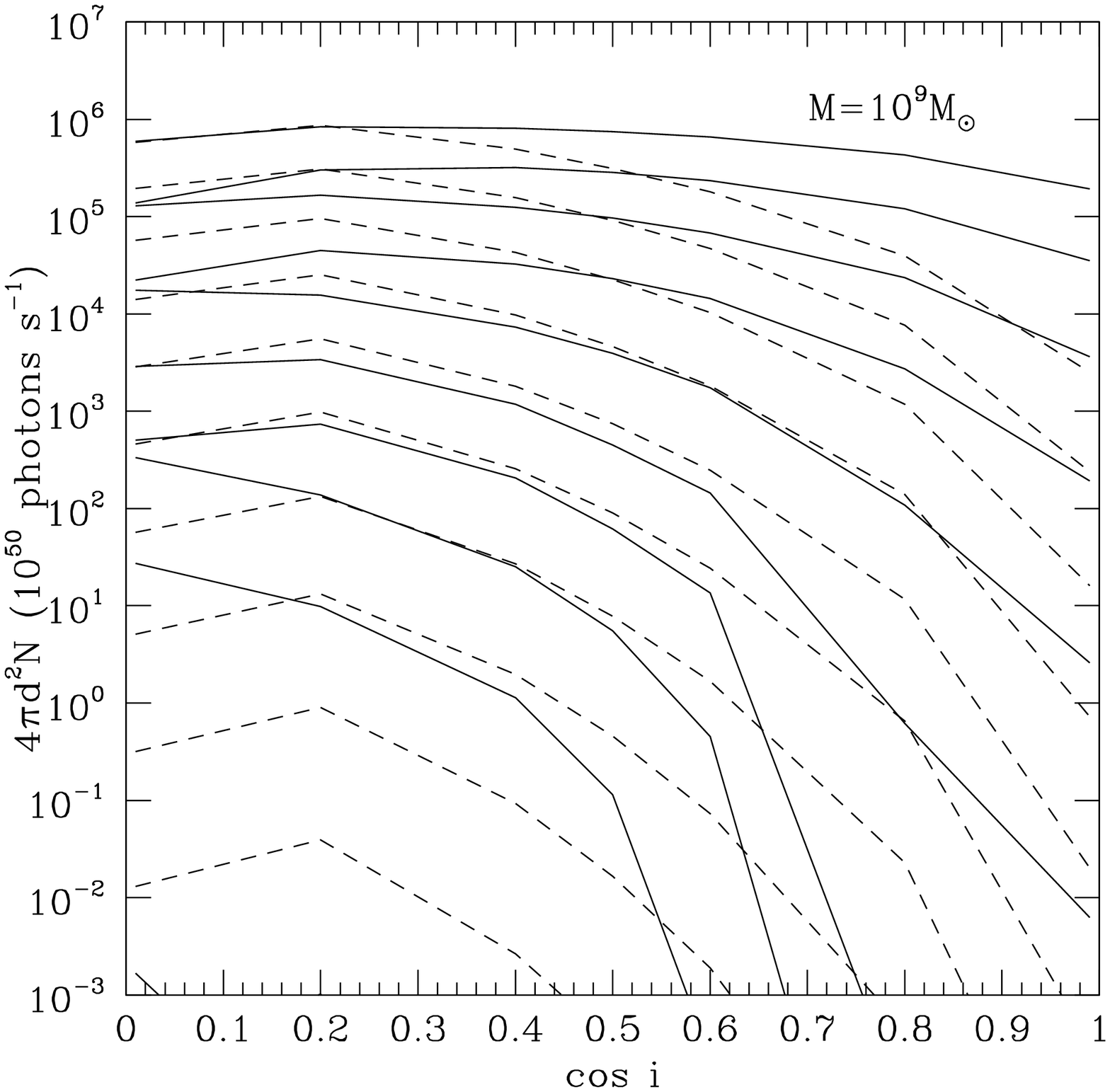,width=3.5in}} 
\noindent{
\scriptsize \addtolength{\baselineskip}{-3pt}
\figcaption{
Same as figure \ref{ION} but for the HeII Lyman continuum.
The lowest accretion rates are off the bottom of the figure.
\label{IONHE}
}
\addtolength{\baselineskip}{3pt}
}

Note that the hydrogen Lyman continuum is limb darkened at high accretion
rates, peaks at intermediate viewing angles at moderate accretion rates, and is
extremely limb brightened at low accretion rates (due to the high Doppler
shifts overcoming the intrinsic absorption edges).  The HeII Lyman continuum
is also limb brightened at all but the highest accretion rates, where it
peaks at intermediate viewing angles.  These intrinsic anisotropies in the
ionizing continuum produced by the disk may have important implications for
photoionization models of the broad line region of AGN (cf. the much
simpler anisotropy model of Netzer 1987) and for the statistics of AGN
samples (e.g. Krolik \& Voit 1998).

\section{Conclusions}

We have presented a grid of AGN disk models for a wide range of
basic parameters, the black hole mass and the mass accretion rate, and for
two values of the viscosity parameter $\alpha$ (0.01 and 0.1), 
and two values of the black hole angular momentum: maximum rotation
Kerr black hole with $a/M=0.998$, and Schwarzschild black hole with
$a/M=0$.

The basic aim of the present study was
to construct a benchmark grid of models, based on simple, ``classical''
approximations, to which all our future, more elaborate models will
be compared.  The most important physical approximations that define
the classical model are the following:\\ 
(i) The energy is generated
by turbulent viscous dissipation, with the vertically-averaged viscosity 
described through the Shakura-Sunyaev parameter $\alpha$.\\
(ii) The vertical dependence of kinematic viscosity is described through
a two-step power law in the column mass.\\ 
(iii) Convection and conduction are neglected.\\ 
(iv) No external irradiation or self-irradiation of the disk is 
considered.\\ 
(v) Electron scattering is treated as coherent (Thomson) scattering,
i.e., the effects of Comptonization are neglected.\\ 
(vi) Thermal opacity and emissivity of H and He only are included here, 
i.e., the effects of metals are neglected.\\ 
(vii) Effects of line opacity are neglected.\\ 
The underlying assumption, not listed here, is that the 1-D approach
is appropriate, i.e., that the disk may be described as a set of
mutually non-interacting, concentric annuli.

What is, however, treated exactly in this study, is the simultaneous
solution of all the structural equations, without making any approximations
concerning the behavior of the radiation intensity. Likewise, no
a priori assumptions about atomic level populations (e.g. LTE) are
made. The local electron temperature and density, mass density, and atomic
level populations, are determined self-consistently with the radiation field.
Once the local spectra of all annuli are computed, the spectrum of the
full disk is found by integrating the emergent intensity
over the disk surface using our relativistic transfer function code.

We can summarize some of the overall spectral features of our benchmark
grid as follows.  Compared to multitemperature blackbody accretion disk
models, our spectra generally have lower fluxes at low frequencies and
higher fluxes at high frequencies.  This difference is amplified further
by relativistic effects, which are strongest for edge-on disks in Kerr 
spacetimes.  Disks with different accretion rates around different mass
black holes do {\it not} exhibit the same spectral energy distribution
even if they have the same effective temperature distribution.  Spectral
slopes in the optical/UV region are significantly redder
than the canonical $F_\nu\propto\nu^{1/3}$ low frequency accretion disk
spectrum.  Non-LTE effects are important in all but the highest density
disks, enhancing the He~II Lyman continuum and generally reducing the
strength of features near the HI Lyman limit.  HI Lyman edge discontinuities
are only present in the cooler, low luminosity disk models.  
High Eddington ratio
models exhibit no discontinuities, but they do show sharp changes in
spectral slope, albeit at wavelengths substantially blue shifted from
912~${\rm\AA}$.  
We stress again that by neglecting convection, the ``cool'' models
($T_{\rm eff} < 9,000$ K may be significantly altered; therefore the 
predicted optical and IR continuum flux (and, in particular, the Balmer edge
region) should be used with caution.  
Finally, our models show substantial wavelength dependent
optical/UV polarization which is parallel to the disk plane, a result which
is likely to be modified by the effects of Faraday rotation and additional
sources of opacity, both of which can suppress this polarization.

In future papers of this series, we will systematically
relax more and more of the classical assumptions listed above.  Among
these, relaxing assumptions (vi) are (vii) is straightforward, since our
computer program TLUSDISK is fully capable of handling these situations. 
The only concern is that generating such models will require much more 
computer time than required for the present models. 
Also, one has to collect a large amount of atomic 
data, but we will profit enormously from already existing collections made
for the purposes of modeling stellar atmospheres, or from data being
included in current photoionization codes.
Relaxing approximation (v) is less straightforward, but we have 
recently solved the problem and implemented Comptonization in TLUSDISK.
Also, approximation (iv) may in principle be easily relaxed by adjusting
the surface boundary conditions of each annulus.

However, relaxing assumptions (i) - (iii) is much more difficult.  Here,
we have to rely on detailed magnetohydrodynamic simulations to guide us in
how to describe convection, and how to choose the most appropriate
parameterization of viscosity.  Nevertheless, even before such simulations
are available, we can investigate phenomenologically the impact of
a dissipation rate that varies with altitude within the disk.  The existence
of disk coronae suggest, for example, that the heating rate may increase
with height, leading to a possible temperature inversion in the upper
layers of disk atmospheres.

Besides this purely theoretical motivation for improving disk models
we will also use the present grid of models to analyze a large
volume of observed quasar spectra. Such a study will bring
interesting results whether or not the models actually fit
the data. If they do, this will be
a strong argument in support of the accretion disk hypothesis, and of the
adequacy of our theoretical description of accretion disks.  If not,
such a study will provide us with important clues as to which aspects
of the theoretical description should be improved, and/or
what other observational constraints will be needed in the future
to settle these questions.

\acknowledgments

We thank Ari Laor for providing us with his spectral models for comparison
with those presented here.  This work was supported in part by NASA grant
NAG5-7075.

\appendix
\section*{Appendix - Density Structure of the Disk}

We consider here some details of the density structure in the case
of a radiation pressure supported disk, without assuming that the
gas pressure is totally negligible.

We write the vertical hydrostatic equilibrium equation as (see Paper II
for details of the formulation)
\begin{equation}
\label{A1}
{d P_{\rm rad} \over d z} + {d P_{\rm gas} \over d z} = - Q\, \rho \, z\, ,
\end{equation}
where $P_{\rm rad}$ and  $P_{\rm gas}$ are the radiation and gas pressure,
respectively, $\rho$ is the mass density, $z$ is the vertical distance from 
the disk midplane, and $Q=(GM/r^3)(C/B)$ [in the notation of Paper II; 
in the notation of Krolik (1999a), $C/B = R_z(r)$], where
$B$, $C$ and $R_z(r)$ are the appropriate relativistic corrections.

The radiation pressure gradient can be written (Paper II; Krolik 1999a)
\begin{equation}
\label{A2}
{d P_{\rm rad} \over d z} = - {\rho \chi_{\rm H}\over c}\, F_{\rm rad}\, ,
\end{equation}
where $F_{\rm rad}$ is the total (frequency-integrated) radiation flux,
and $\chi_{\rm H}$ is the flux-mean opacity. For most applications, the
flux-mean opacity is well approximated by the Rosseland-mean opacity,
$\kappa_{\rm R}$. 
We consider here the case where the local opacity is fully dominated by 
electron scattering, in which case $\kappa_{\rm R}$ is constant and
roughly equal to 0.34. (In fact, it is not exactly constant because it
depends on the degree of ionization - see Paper II; however,
we neglect this small effect here.)
Further, we introduce (Krolik 1999a)
\begin{equation}
\label{A3}
F_{\rm rad}(z)=F_{\rm rad}^0(z)\, f(z) \equiv \sigma_B T_{\rm eff}^4\, f(z)\, ,
\end{equation}
where $F_{\rm rad}^0$ is the total radiation flux at the surface, which
is expressed through the effective temperature.

We express the gas pressure through the sound speed, $c_s$, as
$P_{\rm gas} = c_s^2 \rho$. The sound speed is given by
\begin{equation}
\label{A4}
c_s^2 = {k\over \mu m_{\rm H}}\, {N \over N-n_{\rm e}}\, T,
\end{equation}
where $k$ and $m_{\rm H}$ are the Boltzmann constant and the mass of the
hydrogen atom, respectively; 
$\mu$ is the mean molecular weight (i.e., the mean mass of
a heavy particle per hydrogen atom; in our case of a H-He atmosphere with a
solar helium abundance $\mu=1.4/1.1= 1.27$); $N$ is the total particle
number density, and $n_{\rm e}$ is the electron density. The sound speed
thus depends primarily on the temperature, and partly also on the
degree of ionization
[via the term $N/(N-n_{\rm e})$, which varies from 1 -- for a neutral
medium, to 2.1 -- for a fully ionized solar-composition H-He plasma].

Substituting equations (\ref{A2}), (\ref{A3}), and (\ref{A4}) 
into (\ref{A1}), we obtain
\begin{equation}
\label{A5}
{1\over Q} {d c_s^2\over dz} + {c_s^2\over \rho Q} {d \rho\over dz} = 
H_r f(z) - z
\end{equation}
where we introduce the radiation-pressure scale height, $H_r$, as
\begin{equation}
\label{A6}
H_r \equiv {\kappa_R F_{\rm rad}^0\over c Q} = {\kappa_R \sigma_B T_{\rm eff}^4
\over c Q}\, ,
\end{equation}
which has the meaning of the disk height in the case of negligible gas
pressure (see Paper II and Krolik 1999a). 

We shall consider two cases, (a) the gas pressure is completely
negligible, and (b) the gas pressure is taken into account, although it
is still smaller than the radiation pressure.

In the former case, we take $P_{\rm gas} = 0$, i.e. $c_s=0$, and thus
the l.h.s. of Eq. (\ref{A5}) is zero. We are left with
\begin{equation}
\label{A7}
H_r f(z) - z = 0 \, .
\end{equation}
At the surface, $f(z_0)=1$ regardless of the dissipation law, and therefore
$z_0=H_r$, so the disk height is exactly equal to the radiation pressure
scale height.  However,
the density cancels out exactly because both the gravity force and radiation
force are linearly proportional to density (e.g., Krolik 1999a), so the
density is undetermined by the hydrostatic equilibrium equation. 
In usual treatments of the radiation-pressure-supported disk, Eq. (\ref{A7})
is assumed to hold everywhere in the disk, from
which it follows that $f(z) = z/H_r$. In our approach, however, we assume
that $f$ is a known function of the column mass, $m$. In the case of
constant kinematic viscosity (i.e., the dissipation rate proportional to
density), we have (see Paper II), $f=m/m_0$, where $m_0$ is the column
mass at the midplane, $m_0 = \Sigma/2$. This gives a linear relation between
$z$ and $m$, and since $\rho = -dm/dz$, we obtain $\rho(z) = {\rm const} = 
\rho_0$ in the region of constant dissipation (99\% of the total column
mass of the disk in our models).
In our numerical procedure, we in fact solve Eq. (\ref{A5}) exactly, without
assuming that the left-hand-side is negligible, and with $f$ given as a known
function of $m$ (although not of $z$, because $z$ is one of the state
parameters to be solved as a function of $m$).

To be able to write down a simple analytic solution in the case of 
non-negligible gas pressure, we approximate the function $f(z)$ as
\begin{equation}
\label{A8}
f(z)= \cases{z/H_0\, , \quad {\rm for~} z < H_0\, ,\cr
      1 \, , \quad {\rm for~} z \geq H_0\, .\cr}
\end{equation}
In other words, we assume that the radiation flux linearly increases with $z$
until a certain height $H_0$, and then remains constant.  Such a behavior is
roughly observed in the numerical simulations.  
Since close to the midplane $z/H_0=1-m/m_0$, so that 
$\rho_0 = -(dm/dz)_{z=0} = m_0/H_0$.  Therefore, $H_0= m_0/\rho_0$.
Within the present analytical
model, $H_0$, and thus $\rho_0$ are quantities to be determined by 
the constraint of total column density (see below).
We write 
\begin{equation}
\label{A9}
{c_s^2\over Q} = {c_s^2(z=0)\over Q}\, q(z) \equiv {H_g^2\over 2} \, q(z)\, ,
\end{equation}
where $H_g$ has the meaning of the gas-pressure scale height corresponding to
the midplane conditions (temperature and degree of ionization), and $q(z)$
is a correction parameter that accounts for a dependence of $c_s$ on $z$.
The reason why $H_g$ is called the gas-pressure scale height is that in 
the case of negligible radiation force 
(i.e. $f(z)\rightarrow 0$ everywhere), the
solution of Eq. (\ref{A5}) is given by
\begin{equation}
\label{A10}
\rho(z) \rightarrow \rho_0 \exp[-(z/H_g)^2]\, ,
\end{equation}
where $\rho_0 \equiv \rho(z=0)$ is the density at the midplane.

In the following, we neglect the dependence of sound speed on
height, i.e. we assume $q(z)=1$. One can derive appropriate analytical
expressions even for a more realistic case at the expense of complicating
the analytical formulas considerably, but the present approximation
is satisfactory from the point of view of describing the basic physical 
picture.  A similar analysis was already presented by Hubeny (1990).
Equation (\ref{A5}) then reads, using Eq. (\ref{A9}),
\begin{equation}
\label{A11a}
{1 \over \rho}{d\rho\over dz} = \left( {H_r\over H_0} -1 \right)
{2z \over H_g^2}\, , \quad\quad {\rm for~} z<H_0\,  
\end{equation}
\begin{equation}
\label{A11b}
{1 \over \rho}{d\rho\over dz} =  (H_r-z)\, 
{2 \over H_g^2}\, , \quad\quad {\rm for~} z \geq H_0\, ,
\end{equation}
which has the solution
\begin{equation}
\label{A12}
\rho(z) = \rho_0 \exp\left[-\left(1-{H_r\over H_0}\right)\,
\left({z\over H_g}\right)^2\right]\, , \quad\quad {\rm for~} z<H_0\, ,
\end{equation}
and
\begin{equation}
\label{A13}
\rho(z) = \rho_0
\exp\left[-\left({z-H_r\over H_g}\right)^2\right]\,
\exp\left[-{H_r\over H_g}{H_0-H_r\over H_g}\right]\, ,
\quad\quad {\rm for~} z \geq H_0\, .
\end{equation}
The scale height $H_0$ is now determined from the condition
$\int_0^\infty\rho(z) dz = m_0$.  Substituting Eqs. (\ref{A12}) and
(\ref{A13}), we obtain after some algebra
\begin{equation}
\label{A14}
(2/\sqrt\pi)\, h_0 = \sqrt{h_0/\delta_0}\  
{\rm erf}\left(\sqrt{h_0 \, \delta_0}\right)
+ \exp(-h_r \,\delta_0)\, {\rm erfc}(\delta_0)\, ,
\end{equation}
where
\begin{equation}
h_0 \equiv H_0/H_g\, , \quad h_r \equiv H_r/H_g\, ,
\end{equation}
and
\begin{equation}
\delta_0 = h_0 - h_r \, ,
\end{equation}
where the error function ${\rm erf}(x)$ is defined by 
${\rm erf}(x) \equiv (2/\sqrt\pi) \int_0^x \exp(-t^2)\, dt$,
and the complementary error function by
${\rm erfc}(x) \equiv (2/\sqrt\pi) \int_x^\infty \exp(-t^2)\, dt = 
1 - {\rm erf}(x)$.
Equation (\ref{A14}) is a transcendental equation for $h_0$; however
an approximate solution is found to be
\begin{equation}
h_0 \approx h_r + 1/h_r \, .
\end{equation}

If $h_r \gg 1$ (i.e., $H_r \gg H_g$), one obtains $h_0 \approx h_r$, i.e.,
$H_0 \approx H_r$, and thus $\rho(z) \approx \rho_0$ for $z<H_r$; 
i.e., one recovers the solution for negligible gas pressure. 
If, however, $H_g$ is not completely negligible with respect to $H_r$,
we obtain $H_0 > H_r$, and density falls off exponentially with increasing 
height even in the midplane layer, with a scale height
$H_g/(1-H_r/H_0)^{1/2}$ (see Eq.~\ref{A12}).

From the expression for the respective scale heights we see that
$H_r$ is weakly dependent on $r$ ($H_r \propto D/C$, so it depends on
radial distance only through the relativistic corrections), while
$H_g \propto (T_0/Q)^{1/2}$ where $T_0$ is the temperature at the midplane.
The latter scales as $T_0 \propto m_0^{1/4}\, T_{\rm eff} \propto r^{-3/8}$,
and since $Q \propto r^{-3}$, we obtain
finally $H_g \propto r^{21/16}$, i.e., it increases rapidly with
radial distance. 
For the models displayed in figure \ref{FIGDENS} we have, e.g., at $r/r_g =5$
(which is well in the domain of complete dominance of radiation pressure),
$H_r=2.63 \times 10^{13}$ cm, and $H_g = 5.52\times 10^{11}$ cm, i.e.,
$h_r \approx 48$.  We thus have $H_0 \approx H_r$, which is indeed verified
by the numerical model.
The last (coolest) model displayed there corresponds
to $r/r_g = 90$, and we have 
$H_r=6.42 \times 10^{13}$ cm, and $H_g = 2.67\times 10^{13}$ cm, i.e.,
$h_r \approx 2.4$. The correction $1/h_r$ to $h_0$ is no longer negligible;
we obtain $h_0 \approx 2.8$, i.e., $H_0 \approx 7 \times 10^{13}$ cm
(the exact value following from the numerical model is 
$H_0 = 7.7 \times 10^{13}$ cm)
and the density should show an $\exp[-(z/H)^2]$ decay with $z$ with 
$H \approx 9.3 \times 10^{13}$ cm. This is indeed roughly consistent
with the numerical model.

\clearpage

\end{document}